%% file: main.tex
\documentclass[%
reprint,
superscriptaddress,
nofootinbib,
 amsmath,amssymb,
 aps,
prd,
floatfix,
]{revtex4-1}
\usepackage{amsmath, amssymb}
\usepackage{listings}
\usepackage{graphicx}
\usepackage{dcolumn}
\usepackage{bm}
\usepackage{float,color}
\usepackage{xcolor}
\usepackage[normalem]{ulem}
\usepackage{tabularx}
\usepackage{mathtools} 
\usepackage{multirow}
\usepackage{amssymb}
\usepackage{import}
\usepackage{rotating}
\usepackage{booktabs}
\usepackage{scrextend}
\usepackage{hyperref}
\usepackage{braket}

\begin{document}

\newcommand{\Cardiff}{School of Physics and Astronomy, Cardiff University, Cardiff, CF24 3AA, United Kingdom}
\newcommand{\Soton}{Mathematical Sciences \& STAG Research Centre, University of Southampton, Southampton, SO17 1BJ, United Kingdom}

\title{The High-Mass-Ratio Challenge in Gravitational Waveform Modelling}
\author{Parthapratim Mahapatra}\email{MahapatraP@cardiff.ac.uk}
\affiliation{\Cardiff}
\author{Jonathan E. Thompson}
\affiliation{\Soton}
\author{Edward Fauchon-Jones}
\affiliation{\Cardiff}
\author{Mark Hannam}
\affiliation{\Cardiff}

\date{\today}

\begin{abstract}

Binary black hole (BBH) mergers detected via gravitational waves are addressing key open questions in astrophysics, cosmology, and fundamental physics. Our scientific conclusions rely on extracting accurate source parameters, for which we require accurate signal modelling. It is well known that current BBH waveform models need to be improved for high-mass-ratio, strongly precessing systems, and in this paper we provide a concrete illustration of this issue, showing that the degradation in model performance is substantially more severe than might have been anticipated.
We present numerical relativity (NR) simulations of precessing BBH systems with a mass ratio of 18 and a dimensionless spin of 0.8 on the larger black hole (with the smaller black hole non-spinning), covering five values of spin misalignment.
We assess the accuracy of state-of-the-art waveform models in this region of parameter space by computing the standard mismatch between the models and the NR waveforms. We find that all current waveform models often exhibit significant mismatches ($\gtrsim$0.1), indicating poor performance in this regime.
We also perform limited parameter estimation using a subset of state-of-the-art waveform models, injecting these NR simulations as signals into the three-detector LIGO–Virgo network. In some cases we find errors in mass measurements of over 100\%, dramatically illustrating that substantial improvements are required in existing waveform models. 
The numerical simulations presented here will be valuable for calibrating future BBH waveform models in this region of parameter space.

\end{abstract}

\maketitle

\section{Introduction}\label{sec:intro}

It has now been a decade since the first landmark detection of gravitational waves (GWs) in 2015, an event that inaugurated a new era of observational astronomy~\cite{GW150914}. Over this period, advances in detector sensitivity and network coordination have dramatically expanded the reach of GW science~\cite{GW170817,GW-GRB170817,GW190521,GW190814,GW230529,GW231123,GWTC-4.0}. The 2015-era detector network could observe only a few stellar-mass black hole (BH) mergers per year, whereas today’s detectors routinely observe hundreds, transforming what was once extraordinary into routine~\cite{GWTC1,GWTC2,GWTC3,GWTC-4.0}. 
These observations have already provided key insights into the strong-field nature of gravity~\cite{GW150914-TGR,GW170817-TGR,GWTC3-TGR}, the local expansion rate of the Universe~\cite{GW170817-H0,GWTC-4.0-cosmo}, the physics of ultra-dense matter~\cite{GW170817-EOS}, and the astrophysical population of compact objects~\cite{GWTC-4.0-pop}.
Looking ahead, next-generation ground-based observatories such as the Einstein Telescope~\cite{,ETScience11,Punturo:2010zz} and Cosmic Explorer~\cite{CE:2019iox} are projected to detect thousands of mergers annually, effectively surveying nearly all stellar-origin black-holes in the observable Universe~\cite{Gupta:2023lga}.
Complementing these efforts, the space-based detector LISA will open a new frequency window, granting access to massive black-hole mergers and expanding the multiband potential of GW astronomy~\cite{LISA2017,LISA:2024hlh}. Together, these advances are transforming GW astronomy into an increasingly important component of fundamental physics, astrophysics, and cosmology.

The scientific potential of these observations relies critically on accurate theoretical waveform models of the GW signals as predicted by general relativity (GR)~\cite{LIGOScientific:2016ebw,LIGOScientific:2016vlm,LIGOScientific:2016wkq}. Source properties are inferred by convolving the observed GW data with these model waveforms, within a Bayesian inference framework that systematically identifies the set of model parameters that best reproduce the observed signal~\cite{Bayes1763,Veitch:2014wba}. In this approach, the likelihood of the observed data given a particular waveform model is combined with our prior expectations of the source population to compute a posterior probability distribution, providing both the most likely source parameter values and their associated uncertainties~\cite{Veitch:2014wba,Rover:2006ni,vanderSluys:2008qx}.
Accurate waveform models are essential because they encode the detailed multipolar structure of the signals, ideally including higher-order modes, precession effects, and eccentricity, which are critical to precisely extracting the masses, spins, distances, and orbital properties of the binary~\cite{LIGOScientific:2016ebw,LIGOScientific:2016vlm,LIGOScientific:2016wkq}. Inaccuracies in these models can lead to biased parameter estimates and limit our ability to achieve the science goals of current and future GW detectors, making continued improvements in waveform modeling a key priority for the future of GW astronomy.

The gravitational dynamics of a compact binary can be divided into three stages: inspiral, merger, and ringdown. Due to the complexity of Einstein's equations, there are no exact analytic solutions for the full two-body problem of coalescing binary systems of compact objects. During the early inspiral, Einstein’s equations can be solved perturbatively using the post-Newtonian approximation~\cite{Blanchet:2013haa} and resummation techniques within the effective-one-body approach~\cite{Damour:2009kr}, while key features of the ringdown are described by black hole perturbation theory~\cite{Sasaki:2003xr}. The late inspiral and merger phases, lying between these two regimes, require full solutions of the Einstein equations. After decades of effort to solve the BBH problem in the merger phase, the first complete numerical relativity (NR) simulations spanning a single plunge-orbit, merger, and ringdown were achieved in 2005~\cite{Pretorius:2005gq,Campanelli:2005dd,Baker:2005vv}. Since then, numerous independent NR codes~\cite{Bruegmann:2006ulg,Husa:2007hp,Scheel:2006gg,Hemberger:2012jz,Herrmann:2006ks,Zlochower:2005bj,Sperhake:2006cy,Loffler:2011ay} have been developed, enabling simulations over multiple orbits and accommodating more challenging configurations, including high mass ratios, rapidly spinning black holes, and highly eccentric binaries~\cite{Lousto:2010ut,Sperhake:2011ik,Scheel:2014ina,Hamilton:2023qkv,Ferguson:2023vta,Scheel:2025jct}. 

A BBH system in a quasi-circular orbit is described by eight intrinsic parameters: the redshifted (detector-frame) component masses of the black holes, $m_1$ and $m_2$, and their respective spin vectors, $\vec{S}_1$ and $\vec{S}_2$~\footnote{Throughout, we denote the more massive BH in a binary as the primary and labeled with a “1”.}. In the absence of an intrinsic mass or length scale in vacuum GR, the binary dynamics can be written in terms of the dimensionless combination $fM$, where $f$ is the GW frequency and $M=m_1+m_2$ is the redshifted total mass of the system (with $G=c=1$ throughout this paper). Since the total mass sets the overall frequency scale, it can be conveniently factored out of NR waveforms. 
It is further convenient to introduce the dimensionless spin vector of each black hole as $\vec{\chi}_{i}=\vec{S}_{i}/m_{i}^{2}$ ($i=1,2$). We also define the mass ratio as $q=m_1/m_2\geq1$ by convention.
In addition to these intrinsic parameters, the full gravitational waveform observed at the detector depends on seven extrinsic parameters: the inclination angle $\theta_{\rm JN}$ (angle between the line of sight and the total angular momentum); the luminosity distance $D_{\rm L}$; sky location (right ascension $\alpha$ and declination $\delta$); polarization angle $\psi$; coalescence time, $t_{\rm c}$; and coalescence phase, $\phi_{\rm c}$. For precessing BBHs, the angular parameters are always specified at some reference frequency $f_{\rm ref}$.

Unfortunately, in most cases NR simulations are too computationally expensive to be used directly in data-analysis applications such as parameter estimation (PE) of BBHs from GW observations. Consequently, several fast-to-evaluate approximate models have been developed that generate complete gravitational waveforms that capture all three stages of binary evolution. For instance, the recent GWTC-4.0 catalog of compact binary mergers reported by the LIGO–Virgo–KAGRA (LVK) Collaboration~\cite{GWTC-4.0}, covering the first part of the fourth observing run, relied on waveform models from three main families: {\tt Phenom}~\cite{Ajith:2007kx,Ajith:2009bn}, {\tt SEOBNR}~\cite{Buonanno:2007pf}, and {\tt NRSurrogate}~\cite{Field:2013cfa,Blackman:2015pia}.
The {\tt Phenom} and {\tt SEOBNR} families combine analytic and numerical input to construct complete inspiral–merger–ringdown models, calibrating a theoretically motivated ansatz for the merger–ringdown regime against NR data. In contrast, the {\tt NRSurrogate} models are constructed entirely from NR simulations, interpolating across the available NR data. 
The state-of-the-art waveform models for quasi-circular precessing BBH signals from these families include: {\tt IMRPhenomXPNR} ({\tt XPNR})~\cite{Hamilton:2025xru}, {\tt IMRPhenomTPHM} ({\tt TPHM})~\cite{Estelles:2021gvs}, {\tt SEOBNRv5PHM} ({\tt v5PHM})~\cite{Ramos-Buades:2023ehm}, and {\tt NRSur7dq4}~\cite{Varma:2019csw}.
The {\tt XPNR} model computes the GW strain in the frequency domain, whereas the {\tt TPHM}, {\tt v5PHM}, and {\tt NRSur} models operate directly in the time domain. 

In the {\tt XPNR} model, inspiral double-spin precession is described by evolving the SpinTaylor equations~\cite{Colleoni:2024knd}, and phenomenological fits calibrated to single-spin precessing NR simulations (covering mass ratios up to 8 and dimensionless spins on the larger BH up to 0.8) are applied for the merger and ringdown~\cite{Hamilton:2021pkf}, while the model also incorporates the mode asymmetry of the dominant multipole~\cite{Ghosh:2023mhc}. In contrast, {\tt TPHM} and {\tt v5PHM} are calibrated to NR simulations only for aligned-spin binaries~\cite{Pompili:2023tna,Estelles:2020twz} and model precession by extending post-Newtonian or effective-one-body results~\cite{Khalil:2023kep,Estelles:2020osj}~\footnote{We note that a recent update of the {\tt v5PHM} model incorporates multipole mode asymmetries ({\tt SEOBNRv5PHM\_w/asym})~\cite{Estelles:2025zah}. In this work we use the earlier version, and this choice is expected to have a negligible impact on our conclusions.}. 
All three models include the ($\ell,\, \lvert m \rvert$) = (2, 2), (2, 1), (3, 3), (4, 4) multipolar harmonics, with {\tt XPNR} additionally including the (3, 2) mode, {\tt TPHM} additionally including the (5, 5) mode, and {\tt v5PHM} additionally including the (3, 2), (4, 3), and (5, 5) modes.
The {\tt NRSur} model is fully calibrated to NR waveforms across the BBH parameter space up to spin magnitudes $\chi_{1,2}=0.8$ and mass ratios of 4, with extrapolation extending to $\chi_{1,2}=1$ and mass ratios of 6. By design, it includes all multipoles up to $\ell=4$ as well as precessional features such as mode asymmetry.

There are multiple extensive catalogs of BBH NR simulations, covering different regions of the parameter space~\cite{Mroue:2013xna,Boyle:2019kee,Healy:2017psd,Healy:2019jyf,Healy:2020vre,Healy:2022wdn,Ferguson:2023vta,Hamilton:2023qkv,Scheel:2025jct}. We now have somewhat systematic coverage of quasicircular precessing simulations, extending to mass ratios up to 8 and dimensionless spins up to $\lvert \vec{\chi} \rvert \leq 0.8$~\cite{Hamilton:2023qkv,Scheel:2025jct}. 
NR simulations with mass ratios beyond 8 that are sufficiently long and accurate for gravitational waveform modeling remain sparse. As future GW detectors become more sensitive, they are expected to observe a wider variety of sources, including high mass-ratio binaries and strongly precessing systems. This highlights the need for more systematic NR simulations covering high-mass-ratio binaries with diverse spin orientations.

Here we report NR simulations of highly asymmetric, precessing BBH systems with a mass ratio of 18 and a dimensionless spin magnitude of 0.8 on the larger BH, while the smaller BH is nonspinning. The simulations span five spin misalignment angles, $\theta_{\text{LS}_1}$, defined as the angle between the spin vector of the larger BH and the Newtonian orbital angular momentum: $30^\circ$, $60^\circ$, $90^\circ$, $120^\circ$, and $150^\circ$.
These simulations are primarily intended to facilitate the development of precessing gravitational waveform models calibrated against NR waveforms in the high mass-ratio and strongly precessing regimes.

These new simulations also allow us to demonstrate the critical need for NR calibration of waveform models in the high-$q$, high-precession regime by comparing current models to our new simulations in two ways. The first is to quantify the accuracy of current models by calculating the standard mismatch between the models and NR waveforms.
We find that all existing models generally exhibit substantial mismatches of $\gtrsim 0.1$ (i.e., matches $\lesssim 0.9$), suggesting that they are of poor accuracy in this regime.

Although these mismatches are at least an order of magnitude higher than those typical in the models' calibration region, poor mismatches do not necessarily translate into parameter biases, as discussed in Ref.~\cite{Thompson:2025hhc}, and demonstrated in practice for GW observations in Ref.~\cite{GW231123}. To test the models' performance for parameter recovery~\cite{LIGOScientific:2016ebw,LIGOScientific:2019ysc}, we additionally perform targeted PE using a subset of state-of-the-art waveform models, injecting these NR simulations as signals into the three-detector LIGO–Virgo network. The results confirm that significant improvements in waveform modeling are required to achieve reliable parameter inference for highly asymmetric, strongly precessing binaries.

The outline of the rest of the paper is as follows. In Sec.~\ref{sec:method}, we briefly outline the details of the methods used to assess the accuracy of the waveform models. In Sec.~\ref{sec:nrsim}, we briefly describe the BAM code~\cite{Bruegmann:2006ulg,Husa:2007hp} used to generate the NR simulations of BBH systems presented here, outline the initial data and properties of these simulations, and perform a waveform accuracy analysis to validate them. In Secs.~\ref{sec:accuracy-mismatch} and \ref{sec:accuracy-pe}, we assess the accuracy of current BBH waveform models by comparing them with these NR simulations and highlight the need for improved modeling of highly asymmetric, strongly precessing BBH systems. We present our conclusions and thoughts in Sec.~\ref{sec:conclu}.

\section{Methods for Evaluating Waveform Accuracy}\label{sec:method}

We begin by outlining two complementary methods to quantify waveform accuracy in the context of gravitational wave astronomy: (1) the mismatch, which quantifies the disagreement between waveforms with respect to a detector's noise curve, and (2) parameter estimation, which, given an NR waveform that we assume to be a proxy for the true signal, allows us to quantify how well a set of waveform models will recover individual binary source parameters.

Neither measure is ideal. The mismatch has the advantage of providing a single measure of ``distance'' between two waveforms (see, e.g., Refs.~\cite{Owen:1995tm,Thompson:2025hhc}), but, when evaluated between a fiducial ``true'' waveform and a model, does not tell us how that difference will manifest as loss of signal power, or as parameter biases, or which parameters will be biased and by how much. We can learn these by performing parameter estimation on the true waveform using the model, but parameter estimation is typically far more computationally expensive than computing mismatches, and we can only assess the model's performance for discrete instances of the binary's inclination, polarisation, sky position, etc.  

In this work we use the mismatch to quantify the overall accuracy of our NR simulations, and both the mismatch and parameter estimation to assess the accuracy of current waveform models in the high-mass-ratio, high-precession regime.

\subsection{Mismatches}\label{sec:mismatch}

To quantify the agreement or disagreement between two waveforms --- for example, a model template $h_t$ and an NR signal $h_s$ --- we define the standard noise-weighted inner product as~\cite{Owen:1995tm},
\begin{equation}
    \braket{h_t | h_s} = 4 \text{Re} \int^{f_\text{max}}_{f_\text{min}}
   \frac{\tilde{h}_t \left(f\right) \tilde{h}^{*}_s \left(f\right)}{\tilde{S}_n\left(f\right)} \text{d}f, \label{eq:inner_prod}
\end{equation}
where the tilde denotes a Fourier transform, the waveforms are expressed as functions of frequency $f$, the detector is sensitive in the frequency range $f \in [f_{\rm min}, f_{\rm max} ]$, and $\Tilde{S}_{n}(f)$ is the one-sided power spectral density (PSD) of the detector~\cite{Finn:1992wt,Cutler:1994ys}. The optimal signal-to-noise ratio (SNR) of a GW signal $h$ is given by $\rho=|h|=\sqrt{\braket{h|h}}$.
A commonly used measure to evaluate the accuracy of the waveform model is the {\it match}, defined as the noise-weighted inner product between two normalized waveforms, maximized over a subset of template parameters $\theta_{\text{opt}}$ described below:
\begin{equation} \label{eqn: match_def}
   \mathcal{M}\left(h_t,h_s\right) =  \max_{\theta_\text{opt}} \, \frac{\Braket{h_t|h_s}} { \left|h_t\right| \left|h_s\right|}.
\end{equation}
The match is equal to unity when the two waveforms are identical, aside from an overall amplitude scaling. To quantify the difference between two waveforms, it is convenient to define the {\it mismatch},
\begin{equation}
\label{eq:mismatch}
\mathfrak{M}\left(h_t,h_s\right)=1-\mathcal{M}\left(h_t,h_s\right).
\end{equation}
A lower mismatch indicates better agreement between the two waveforms. In GW data analysis, assuming that the statistical likelihood is approximately Gaussian (which is valid in the high-SNR limit), waveform modelling systematics will not bias the PE if the model’s mismatch with the true signal (typically represented by NR simulations that serve as proxies for the true signals) is smaller than $\chi_{k}^{2}(1-p)/2 \rho^{2}$, where $\chi_{k}^{2}(1-p)$ is the chi-square value at probability $p$ for $k$ degrees of freedom~\cite{Lindblom:2008cm,McWilliams:2010eq,Baird:2012cu}. For single-parameter measurements (i.e., $k=1$), this criterion for the 90\% credible interval reduces to $\mathfrak{M}\le1.35/\rho^{2}$.
As an example, for a signal with SNR $\rho = 30$, the mismatch between the waveform model and the NR simulation must be $\mathfrak{M} \lesssim 1.5 \times 10^{-3}$ to prevent the true parameter from lying outside the 90\% credible interval of the marginalized 1D posterior distribution. However, this is only a conservative criterion: a model will not report biased parameters if it meets this criterion, but failing the criterion does not necessarily imply a bias. For a detailed discussion of the proper use and interpretation of this mismatch criterion, we refer interested readers to Ref.~\cite{Thompson:2025hhc}.

We now explain the set of parameters $\theta_{\text{opt}}$ that are optimized to obtain the maximized match. Shifting $t_{\rm c}$ simply moves the binary’s coalescence time, while adjusting $\phi_{\rm c}$ changes only the binary’s initial orbital orientation. For non-precessing waveforms that include only the quadrupolar modes,  $\psi$ is degenerate with $\phi_{\rm c}$. In addition, the inclination, luminosity distance, and the sky-position angles are mutually degenerate, since variations in these parameters merely rescale the overall amplitude of the waveform. Thus, the waveform depends on these parameters only through an overall phase and an overall amplitude. 

In the precessing case, the source’s orientation relative to the detector evolves as the orbit precesses. Consequently, the symmetries present in the non-precessing case no longer apply; both the inclination and the polarization angle must be explicitly considered.
We follow the procedure for computing the optimized match for precessing BBH systems outlined in Refs.~\cite{Schmidt:2014iyl, Harry:2016ijz} and implemented in Ref.~\cite{Hamilton:2021pkf} (see Sec.~XI~A therein for further details).
For signals from non-precessing BBH systems that consist only of the dominant $(\ell, |m|) = (2, 2)$ modes, an analytical expression can be derived for the match maximized over $D_{\rm L}$, $\phi_{\rm c}$, $\psi$, $\alpha$, and $\delta$ of the model template $h_t$.
Ref.~\cite{Harry:2016ijz} provides an analytical expression for the SNR maximized over these parameters (see Eq.~(27) and Secs.~III and IV for a detailed derivation of Eq.~(27)). Dividing the maximized SNR by the signal norm, i.e., $\braket{h_s|h_s}$, yields the match maximized over $D_{\rm L}$, $\phi_{\rm c}$, $\psi$, $\alpha$, and $\delta$ of the model template $h_t$. Since the coalescence time $t_{\rm c}$ parametrizes time translations of $h_t$, the match can be further maximized over $t_{\rm c}$ in a computationally efficient manner using an inverse Fast Fourier Transform routine~\cite{Balasubramanian:1995bm, Ohme:2012cba}~\footnote{We note that, when using the inverse Fast Fourier Transform routine for timeshift optimization, the resolution of the discrete timestep is increased by padding the frequency-domain data. This is particularly important for signals that differ only slightly in the linear-in-frequency contributions to their phases~\cite{Ohme:2012cba, Ajith:2012az}.}. It should be noted that all parameters are defined at the reference frequency $f_{\text{ref}}$.

This procedure can be extended to templates that include higher-order modes and precession, allowing estimation of the match maximized over $D_{\rm L}$, $\psi$, $\alpha$, $\delta$, and $t_{\rm c}$ of the model template $h_t$. For templates with higher-order modes, each $(\ell,m)$ mode carries an $m \phi_{\rm c}$ dependency; thus, we further optimize the match numerically over $\phi_{\rm c}$.
We also numerically optimise the match over rotations of the in-plane spin components of the template at the reference frequency, following Refs.~\cite{Pratten:2020ceb, Hamilton:2021pkf}.
In order to perform the numerical optimizations, we use the Nelder-Mead algorithm as implemented in the {\tt SciPy} Python package~\cite{2020SciPy-NMeth}. While computing the optimized match, we fix the masses and other spin parameters (leaving only the first Euler angle, or precession angle, free), as well as the inclination angle $\theta_{\rm LN}$ (angle between the line of sight and the orbital angular momentum at the reference frequency $f_{\rm ref}$) to be the same for both the template $h_t$ and the signal $h_s$.

Here, we adopt the {\tt aLIGOZeroDetHighPower} PSD of advanced LIGO~\cite{ALIGO_det} with $f_{\text{min}}=\text{max}\{20\, \text{Hz},\, 1.35 \times f_{\text{min}}^{\text{NR}}\}$ and $f_{\text{max}}=2048$ Hz, where $f_{\text{min}}^{\text{NR}}$~\footnote{$f_{\text{min}}^{\text{NR}}$ is conventionally the same as the reference GW frequency $f_{\text{ref}}$, at which the binary parameters are defined.} is frequency of the (2, 2)-mode in Hz at the beginning of the NR waveform, measured in a frame whose $z$-axis is instantaneously aligned with the orbital angular momentum. We compute the optimized match over $D_{\rm L}$, $\phi_{\rm c}$, $\psi$, $\alpha$, and $\delta$ of the model template for optimally oriented signals (i.e., $\alpha=\delta=0$ for $h_s$) in a single detector.
We consider representative values of the redshifted total mass $M$.
For each mass, we adopt thirteen equally spaced inclination angles from $\theta_{\rm LN} = 0$ to $\pi$. For each combination of $(M, \theta_{\rm LN})$, we evaluate the optimized match over a grid of signal phase and polarization values, $\phi_{\rm c, \mathrm{s}}$ and $\psi_{\mathrm{s}}$: eight equally spaced $\phi_{\rm c, \mathrm{s}}$ values from $0$ to $2\pi$ and thirteen equally spaced $\psi_{\mathrm{s}}$ values from $0$ to $\pi$. For each combination, we record the maximum and minimum optimized match values.

The number of GW cycles in the frequency range of 20\mbox{--}2048\,Hz can vary significantly between different binary configurations, even for a fixed total mass $M$. The mismatch depends strongly on the total cycles, since in general it is easier to optimise Eq.~(\ref{eqn: match_def}) for a less complex signal; this point is illustrated in Ref.~\cite{Mitman:2025tmj}. This should be borne in mind when attempting to compare mismatches between our five configurations, where, for a binary total mass of $150 M_{\odot}$, the number of GW cylces in band varies between $\sim$30 ($\theta_{\rm LS_1} = 30^\circ$) and $\sim$5 ($\theta_{\rm LS_1} = 150^\circ$). 

Since lower matches are typically associated with edge-on systems that have reduced SNR and consequently a lower likelihood of source detection, we then follow Refs.~\cite{Schmidt:2014iyl, Khan:2018fmp, Pratten:2020ceb, Hamilton:2021pkf} and also evaluate the SNR-weighted match in this work. The SNR-weighted match is obtained by averaging the optimized match over different signal phases and polarizations, with each contribution weighted by the corresponding SNR, thereby accounting for the detectable volume,
\begin{equation}
\label{eq:weightedMatch}
\mathcal{M}_\mathrm{w} =
\left(
\frac{
	\sum\limits_{\psi_\mathrm{s},\,\phi_{\rm c, \mathrm{s}}} 
	\mathcal{M}^3 \, \langle h_\mathrm{s} | h_\mathrm{s} \rangle^{3/2}
}{
	\sum\limits_{\psi_\mathrm{s},\,\phi_{\rm c, \mathrm{s}}} 
	\langle h_\mathrm{s} | h_\mathrm{s} \rangle^{3/2}
}
\right)^{1/3},
\end{equation}
where the summation runs over different values of the coalescence phase and polarization of the source, i.e., in our case, of the NR waveform.
The SNR-weighted mismatch is defined as $\mathfrak{M}_\mathrm{w}=1-\mathcal{M}_\mathrm{w}$. We note that an identical mismatch calculation was also employed in the waveform systematics study of the heaviest BBH merger event, GW231123, reported to date by the LVK Collaboration~\cite{GW231123}.

\subsection{Parameter Estimation}\label{sec:pe}
We further assess model accuracy by investigating the parameter biases incurred during Bayesian parameter estimation. This offers a direct view of how waveform inaccuracies translate into biases in the inferred source properties.
Bayesian inference is the process of estimating the parameters $\vec{\theta}$ that characterize a signal, given a model or hypothesis $h$ for the signal (here, a waveform model for the GW signal in GR) and observed data $d$ (here, GW data) corresponding to that signal. Consider the time series $d$ measured by the ground-based GW interferometers. Under the hypothesis that a GW signal exists in the data, the data can be modeled as $d=s+n$, with $s$ being the true signal and $n$ the detector noise. The \emph{posterior probability distribution} on model parameters $\vec{\theta}$, can be obtained via Bayes’ theorem
\begin{equation}
    \label{eq:Bayes_theorem}
    p(\vec{\theta} | d, h) = 
\frac{\mathcal{L}(d | \vec{\theta}, h) \, \pi(\vec{\theta} | h)}{\mathcal{Z}(d | h)},
\end{equation}
where, $\pi(\vec{\theta} | h)$ is the \emph{prior probability} distribution of  $\vec{\theta}$, $\mathcal{L}(d | \vec{\theta}, h)$ is the \emph{likelihood function} and $\mathcal{Z}(d | h)\equiv\int \mathcal{L}(d | \vec{\theta}, h) \, \pi(\vec{\theta} | h) d\vec{\theta}$ is the \emph{Bayesian evidence} in favor of the model $h$. Under the assumption that the detector noise $n$ is wide-sense stationary and Gaussian, the likelihood function is proportional to~\cite{Finn:1992wt,Cutler:1994ys}
\begin{equation}
    \label{eq:likelihood}
    \mathcal{L}(d | \vec{\theta}, h) \propto \text{exp} \bigg\{-\frac{1}{2} \braket{ d-h(\vec{\theta}) | d-h(\vec{\theta})}\bigg\}.
\end{equation}

For quasi-circular, precessing BBHs, evaluating the posterior distribution in GW astronomy involves a 15-dimensional integral, making direct calculation challenging. A common approach to reduce this complexity is to marginalize analytically over selected parameters~\cite{Veitch2013, Farr2014, Singer:2015ema,Thrane2019PASA}; in this work, we adopt this approach for the luminosity distance~\cite{Singer:2015ema, Singer:2016eax}. Due to the high dimensionality of the evidence integral, stochastic sampling methods are commonly used to generate samples from the posterior distribution. A variety of software packages exist for performing Bayesian inference in GW astronomy, many of which implement the nested sampling algorithm. Here, we use the Bayesian parameter inference package {\tt Bilby}~\citep{bilby_paper,bilby_pipe_paper} with the {\tt Dynesty}~\citep{dynesty} sampler, interfaced via the {\tt Bilby-pipe} wrapper~\cite{bilby_pipe_paper}, to generate discrete samples for the probability distribution function $p(\vec{\theta} | d, h)$. 

Our sampler setup closely follows that used by the LVK Collaboration in their production analyses~\cite{GWTC-4.0,GWTC-4.0-method}, employing the \texttt{rwalk} algorithm implemented in \texttt{Bilby}. All analyses use 1000 live points and an average of 60 accepted steps per Markov Chain Monte Carlo iteration. We also adopt the prior probability distribution $\pi(\vec{\theta} | h)$ following the standard setup used in the publicly released LVK analysis~\cite{GWTC-4.0,GWTC-4.0-method}. Uniform priors are assigned for the redshifted component masses and spin magnitudes of the individual BHs, while the spin orientations and other angular parameters are sampled isotropically. The luminosity distance $D_{\rm L}$ prior corresponds to sources uniformly distributed in comoving volume and source-frame time. These choices provide broad, uninformative priors that encompass the relevant parameter space while preserving computational efficiency. Overall, these chosen priors are broad and uninformative, ensuring that the posterior support lies well within the prior bounds while maintaining computational efficiency.

We first generate a synthetic GW-signal injection from the reported NR simulations following the LVK NR injection infrastructure \cite{Schmidt:2017btt}, using a fiducial redshifted total mass and the extrinsic parameters specified below. All injections assume the “zero-noise” approximation in a three-detector LIGO–Virgo network~\cite{Acernese_2015,Aasi_2015,KAGRA:2013rdx}, employing the predicted noise power spectral densities at design sensitivity: {\tt aLIGOZeroDetHighPower} for the LIGO detectors~\cite{ALIGO_det} and {\tt AdvVirgo} for the Virgo detector~\cite{AVirgo_det}. In this setup, $d$ corresponds to the time-series strain data generated from the NR simulations. 
The zero-noise approximation enables us to isolate potential systematic biases and avoid statistical fluctuations due to specific noise realizations.
For all NR injection analyses reported here, we adopt the following parameter values for the source’s sky location and polarization: $\alpha=240^\circ$, $\delta=45^\circ$, and $\psi=60^\circ$. We also choose an optimal network SNR of $\sim$50 for the injected NR signals, adjusting the luminosity distance for each injection accordingly.

We then perform Bayesian parameter estimation using the three state-of-the-art waveform models {\tt XPNR}, {\tt TPHM}, and {\tt v5PHM}.
A lower cut-off frequency of 20 Hz is adopted when evaluating the likelihood integral in Eq.~\ref{eq:likelihood}, while the upper cut-off frequency varies between systems, primarily depending on the redshifted total mass. These frequency limits are selected to retain the signal content while avoiding unnecessary computational overhead from excessively high frequency cut-offs. We set both the injected NR signal and the recovery template waveform to start at $f_{\text{min}}^{\text{NR}}$, to avoid missing frequency content from higher-order multipoles~\cite{Ursell:2025ufb}. This is particularly important for Bayesian PE analyses that use frequency-domain waveform approximants. The value of $f_{\text{min}}^{\text{NR}}$ varies across NR simulations and is determined by the characteristics of the NR waveform and the redshifted total mass.
Other analysis settings—such as the signal duration and sampling frequency—are consistent with standard practices used in LVK studies~\cite{GWTC-4.0,GWTC-4.0-method}, and are adjusted for each system based on the physical characteristics of the corresponding injection. The sampler configuration has been described previously and remains the same across all analyses. Lastly, we note that we usually generate synthetic GW signals from NR simulations using all the available modes (i.e., $\ell \leq 4$ modes) and perform the PE recovery using all the available modes in the waveform approximant, unless otherwise specified.

We compare the true injected parameters $\vec{\theta}_{\rm Inj}$ of the NR signal with the posterior distribution inferred from the PE analysis to assess how well the model captures the underlying signal. 
In the zero-noise approximation, the inferred posterior distribution should peak at the true parameters $\vec{\theta}_{\rm Inj}$ when the model perfectly describes the NR signal i.e., $s(\vec{\theta}_{\rm Inj})=h(\vec{\theta}_{\rm Inj})$~\footnote{While sampling error and the finite number of posterior samples typically introduce only negligible shifts, parameter degeneracies can in principle lead to much larger offsets, though they are not expected to significantly affect the results reported here.
}. 
If the model does not faithfully represent the underlying NR signal, the resulting posterior distribution can exhibit systematic biases, i.e., the posterior may peak at values different from $\vec{\theta}_{\rm Inj}$.
In GW astronomy, it is common practice to regard a posterior distribution as biased when the marginalized one-dimensional distribution excludes the true parameter value from its 90\% credible interval.

\section{NR Simulations}\label{sec:nrsim}

We present a set of NR simulations of high-mass-ratio, high-spin, highly precessing BBHs, motivated by the lack of simulations in this region of parameter space. An earlier set of simulations of non-precessing binaries at mass-ratio 18 was used a decade ago to calibrate the aligned-spin BBH waveform model {\tt IMRPhenomD}~\cite{Khan:2015jqa,Husa:2015iqa}, as well as the aligned-spin sector of successor {\tt Phenom} models~\cite{London:2017bcn}. The most recent SXS catalog contains precessing binaries up to mass-ratio 15, but there is only one configuration with a primary spin of 0.8, all others are below 0.5~\cite{Scheel:2025jct}.
Simulations at larger mass ratios are also available (for example, in the RIT catalog~\cite{Lousto:2020tnb}) but are restricted to aligned-spin configurations.
A set of highly precessing, high-mass-ratio simulations enables an initial assessment of how well current BBH waveform models extrapolate into this region of parameter space and provides guidance for first attempts to extend modelling efforts for precessing binaries to higher mass ratios.

Our ability to cover the high-$q$ region in NR simulations is limited by the high computational cost of simulations in this part of parameter space.
The finest length scale that must be resolved in our simulations, which also determines the timestep, is dictated by the smallest BH, while the overall time scale of the simulation is determined by the total mass of the binary; this introduces an approximate factor of $q$ in the scaling of the computational cost. Since the time to merger from a given start frequency also scales as $q$, the computational cost is expected to scale at least as $q^2$; see, for example, Ref.~\cite{LISAConsortiumWaveformWorkingGroup:2023arg}. In addition, the code we use, BAM~\cite{Bruegmann:2006ulg,Husa:2007hp}, is based on Cartesian boxes, which implies a prohibitive computational cost in maintaining sufficient angular resolution for wave extraction as we extract further from the source. At the extraction radii that are feasible in our simulations $\sim$$100\,M$, we risk significant errors in both the amplitude and phasing of sub-dominant modes. In an earlier set of precessing-binary simulations, which extended up to mass-ratio 8, we found that the wave extraction error was the dominant error source~\cite{Hamilton:2023qkv}. For both of these reasons, we consider only a limited set of simulations and prioritize the most strongly precessing configurations.

\begin{table*}[htbp]
    \input{metadata-tabular-q18}
    \caption{Initial-data parameters, binary properties, and remnant properties of each NR simulation. $M\omega_{\rm orb}$ is the binary's initial orbital frequency, $t_{\rm M}$ is the time to merger, and $N_{\rm orb}$ is the number of orbits to merger. See text for more details. }
 \label{table:metadata}
\end{table*}

We simulate five configurations at mass-ratio 18, with a spin of 0.8 on the larger BH (the smaller BH is non-spinnning), and, as in Ref.~\cite{Hamilton:2023qkv}, a set of five spin misalignment angles. The simulations are $\sim$3000$M$ in length; these are roughly 50\% longer than the earlier BAM simulations used for waveform modeling~
\cite{Husa:2015iqa,Hamilton:2023qkv}, in order to capture as large a frequency range as feasible with the computational resources available to us. From a given starting frequency, the total inspiral time scales as approximately $1/\eta$~\cite{Blanchet2024,LISAConsortiumWaveformWorkingGroup:2023arg}, where $\eta = \tfrac{q}{(1+q)^{2}}$ is the symmetric mass-ratio. For a binary with $q=18$, $\eta \approx 0.05$, so the total inspiral time will be roughly five times that of a comparable equal-mass system, with $\eta = 0.25$ if starting at the same orbital frequency. For this reason, although our simulations are longer than in the earlier BAM cataloq, the starting frequencies are higher. 

We construct our initial data, find low-eccentricity parameters, and set up the computational grids following the same procedures and criteria as in Ref.~\cite{Hamilton:2023qkv}. Specifically, we generate initial data by using the Bowen-York analytic solution to the momentum constraints for two boosted, spinning black holes~\cite{Bowen:1980yu} and solve the Hamiltonian constraint in the puncture construction~\cite{Brandt:1997tf} with a pseudo-spectral solver~\cite{Ansorg:2004ds}.
We evolve the data using the standard moving-puncture treatment~\cite{Baker:2005vv,Campanelli:2005dd} of the BSSN decomposition~\cite{Shibata:1995we,Baumgarte:1998te} of the 3+1 Einstein equations~\cite{York:1978gql}. We use sixth-order-accurate spatial finite differencing in the bulk~\cite{Husa:2007hp}, and fourth-order Runge-Kutta time integration with a Courant factor of 0.25. To adequately resolve each BH, the smallest box following each BH should have a width between 1.2 and 1.5 times the maximum coordinate diameter of the apparent horizon. We also require sufficient resolution to resolve the $(4,4)$ multipole in the ringdown, with a grid spacing of $\leq1/(20 f_{\rm RD})$, where $f_{\rm RD}$ is the ringdown frequency of the dominant $(2,2)$ mode. Within these specifications, the basic indicator of numerical resolution is the number of points in each direction on the finest refinement level. For the simulations reported in Ref.~\cite{Hamilton:2023qkv}, the standard box size on the finest level was $96^3$ points, while here we have used $120^3$ points in an attempt to achieve comparable phase accuracy over a longer simulation. We discuss accuracy further in Sec.~\ref{sec:nraccuracy}; we find that the computational cost to achieve accuracy comparable to earlier simulations is prohibitive, although our new simulations are accurate enough for initial modeling purposes and for a preliminary assessment of the accuracy of current waveform models.

Selected properties of the five NR simulations are listed in Table~\ref{table:metadata}. All configurations have mass-ratio $q=18$, primary spin magnitude $\chi_1 = 0.8$ and no spin on the secondary. The  primary's spin is tilted with respect to the Newtonian orbital angular momentum (i.e., the normal to the orbital plane) by $\theta_{\rm LS_1}$. We report the standard effective spin parameters, the effective inspiral spin $\chi_{\mathrm{eff}} \equiv (\chi_{1}\cos\theta_{\rm LS_1}+q\chi_{2}\cos\theta_{\rm LS_2})/(1+q)$~\cite{Ajith:2009bn} and $\chi_p\equiv {\rm max}(\chi_{1}\sin\theta_{\rm LS_1}, \, q \tfrac{3+4q}{4+3q} \chi_{2}\sin\theta_{\rm LS_2})$~\cite{Schmidt2015PhRvD}, as calculated from the initial parameters of the simulation.
The separation at the beginning of each simulation is $D/M$. 
The eccentricity $e$ is estimated over the interval $[200, 1000]M$ of the simulation using the method described in \cite{Husa:2007rh}. The orbital frequency $M\omega_\mathrm{orb}$ is derived from the puncture dynamics at the relaxed time $t=50M$. The time at merger, $t_M$, is defined as the duration from the relaxed time until the separation between the two punctures falls to $0.5M$.
The number of orbits, $N_{\mathrm{orb}}$, is calculated from the relaxed time, where $M\omega_\mathrm{orb}$ is reported up to $t_M$. The mass and spin of the remnant BH, $M_f$ and $\chi_{f}$, are calculated from the apparent horizon of the final black hole. The radiated linear momentum is calculated as described in Ref.~\cite{Bruegmann:2006ulg}, and and the recoil (or kick) velocity $V_{\rm kick}$ of the remnant BH is obtained by integrating the radiated linear momentum from the relaxed time until the end of the simulation.  

\begin{figure}[htbp]
    \centering
\includegraphics[width=0.485\textwidth]{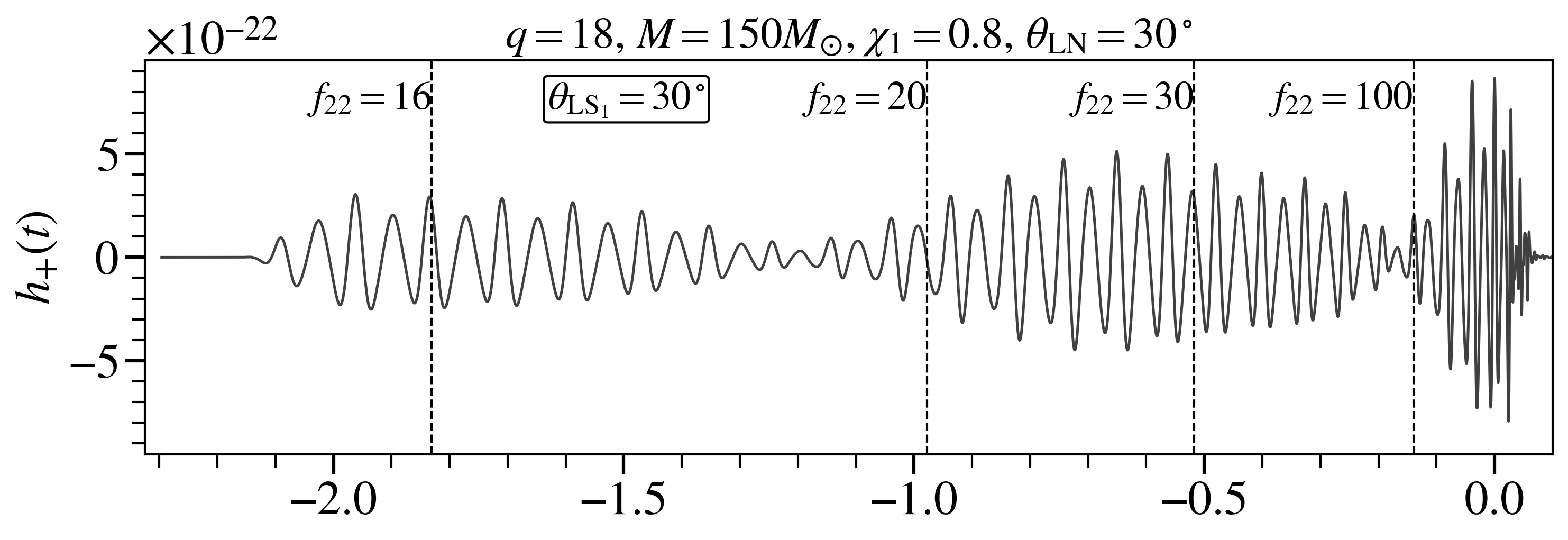}\\
\includegraphics[width=0.485\textwidth]{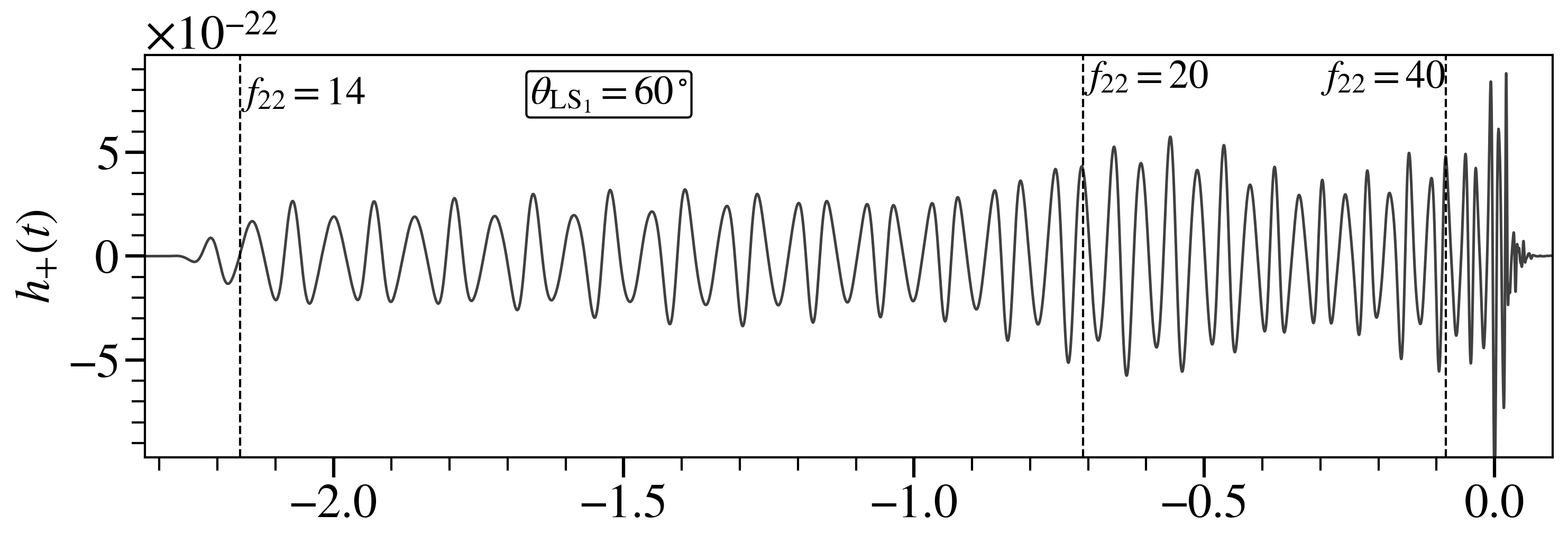}\\
\includegraphics[width=0.485\textwidth]{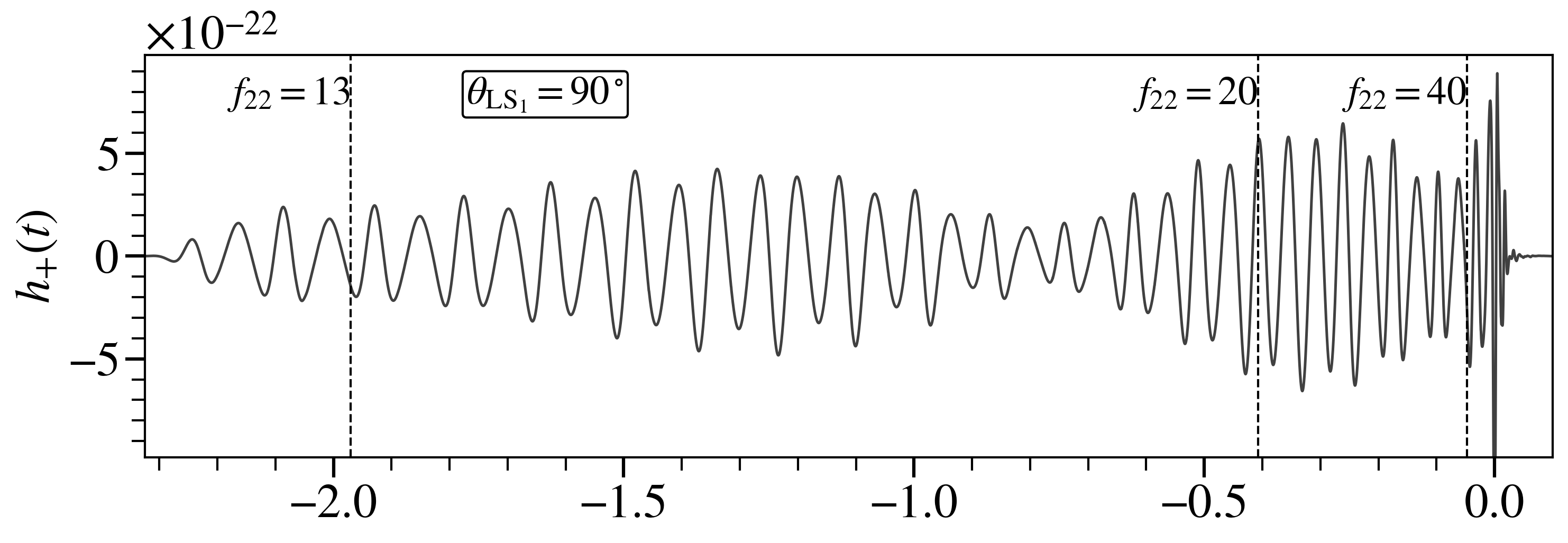}\\
\includegraphics[width=0.485\textwidth]{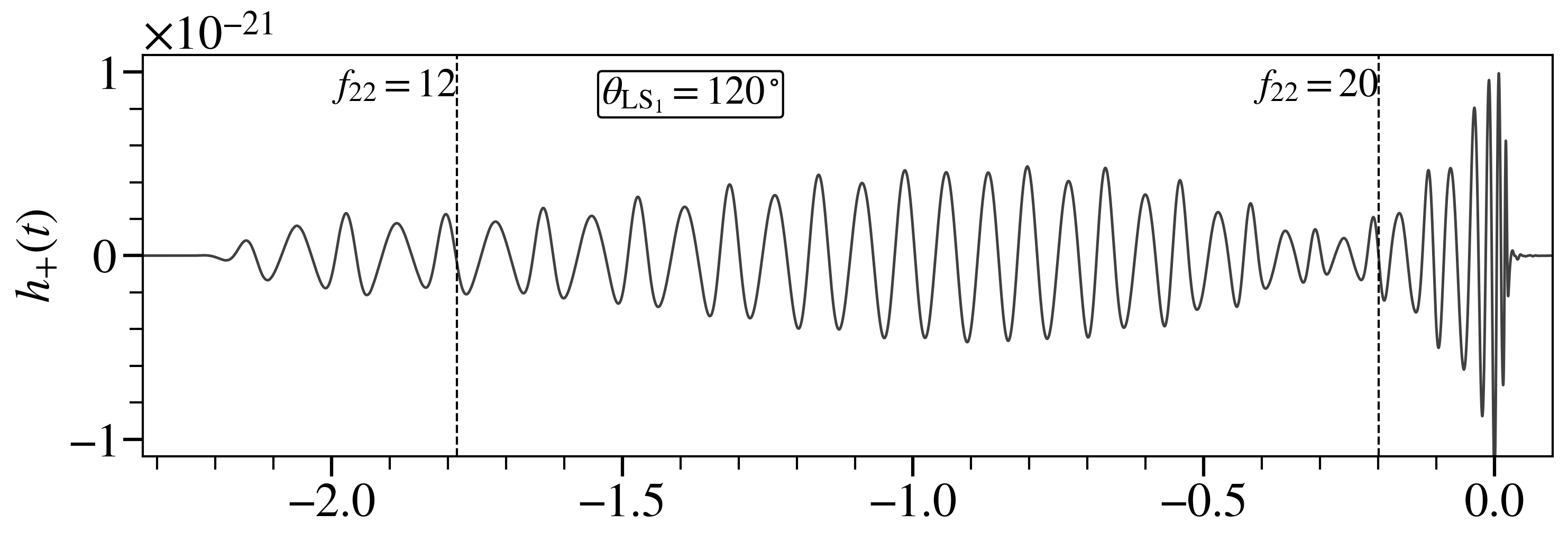}\\
\includegraphics[width=0.485\textwidth]{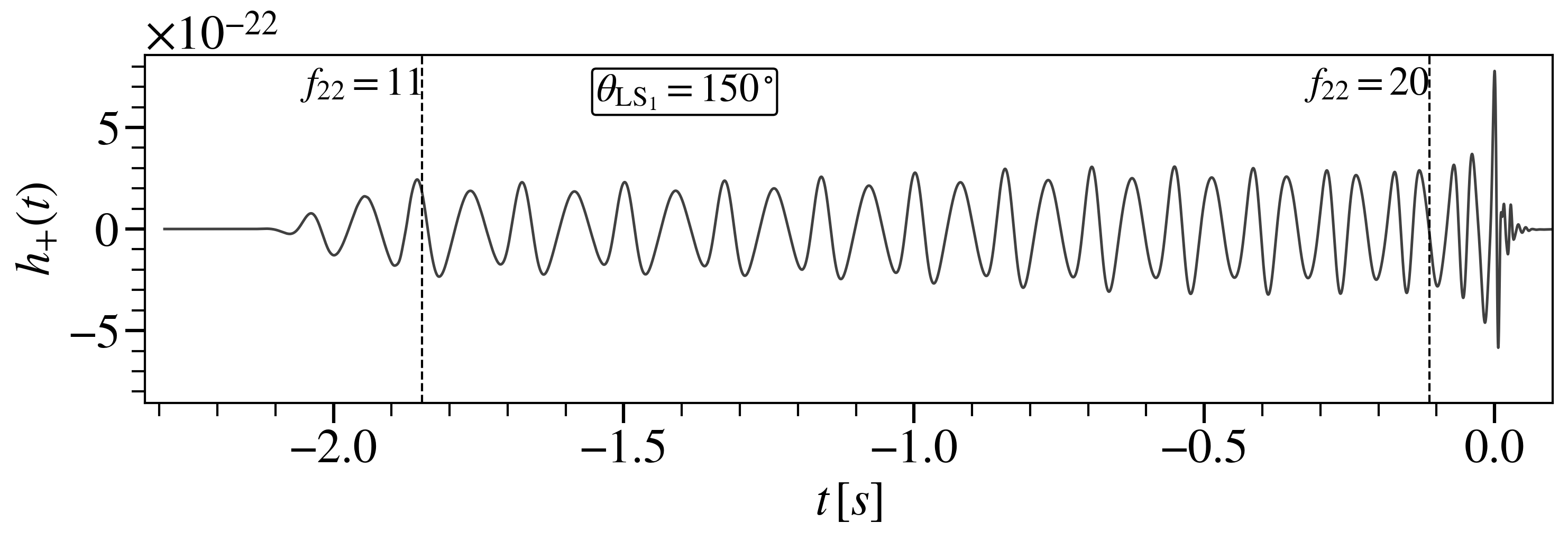}
\caption{The plus GW polarization of each of our NR simulations, with redshifted total mass $150 M_{\odot}$, $\theta_{\mathrm{LN}}=30^\circ$, and $D_L=300$ Mpc. The five panels show each choice of spin misalignment: $\theta_{\rm LS_1}=30^\circ, 60^\circ,\, 90^\circ,\, 120^\circ,\, 150^\circ$. The waveforms have been tapered to remove junk radiation.}
    \label{fig:NR_sim}
\end{figure}

The GW content of the binary system is extracted at finite radius from the source using the Newman–Penrose scalar $\psi_4$~\cite{Newman:1961qr}, following the procedures outlined in Ref.~\cite{Baker:2001sf}. The observable GW strain $h$ is obtained by performing a double time integration of $\psi_4$~\footnote{In practice, when required for applications such as waveform modelling, we compute it in the frequency domain by dividing the Fourier transform of $\psi_4$ by $\omega^2$, where $\omega$ is the angular frequency~\cite{Reisswig:2010di}.}. Here, we have extracted up to $\ell=4$ multipole moments of the gravitational-wave strain; the memory modes (i.e., multipole moments with $m=0$) are not included. Fig.~\ref{fig:NR_sim} shows plus polarization of the gravitational-wave strain in the time domain from the five reported NR simulations for an illustrative binary with redshifted total mass $150 M_{\odot}$, $\theta_{\mathrm{LN}}=30^\circ$, and $D_L=300$ Mpc (polarization angle set to zero). The reported waveforms reveal strikingly rich modulations, arising from the combined effects of spin precession and higher-harmonic contributions, highlighting the intricate structure of high mass-ratio precessing BBH signals.

As noted earlier, the total number of inspiral cycles from a given starting frequency scales approximately as $\sim 1/\eta$~\cite{Arun:2004hn,Kidder:1995zr}. For a binary with mass ratio $q=18$, this results in roughly five times more cycles than a nearly equal-mass system from the same starting frequency. 
For example, GW190521, which was a nearly equal-mass binary with a total redshifted mass of $150 M_{\odot}$ and some evidence of spin-induced orbital precession, had only a few cycles in the LIGO–Virgo band~\cite{GW190521}. In contrast, at 150\,$M_\odot$ our {\tt CF\_83} simulation includes almost ten GW cycles from 20\,Hz to merger; see the third panel of Fig.~\ref{fig:NR_sim}. In this configuration the spin lies entirely in the orbital plane, so there is no aligned-spin component and we expect the inspiral rate to be similar to a  non-spinning binary with the same masses. 

These configurations also clearly illustrate the ``orbital hang-up'' effect~\cite{Kidder:1992fr, Kidder:1995zr}: when BH spins are co-aligned with the orbital angular momentum, the binary inspirals for longer before merger, producing more GW cycles. Anti-aligned spins, in contrast, lead to faster mergers with fewer cycles. This effect is evident in our simulations. Considering the instant of binary separation when $f_{22} =20$ Hz in all five cases shown in Fig.~\ref{fig:NR_sim}, we find that the number of cycles to merger decreases significantly as the spin misalignment angle with the orbital angular momentum increases (top to bottom panels in Fig.~\ref{fig:NR_sim}), with the inspiral shortening due to the orbital hang-up effect.

\begin{figure}[hb]
    \centering
    \includegraphics[width=0.975\columnwidth]{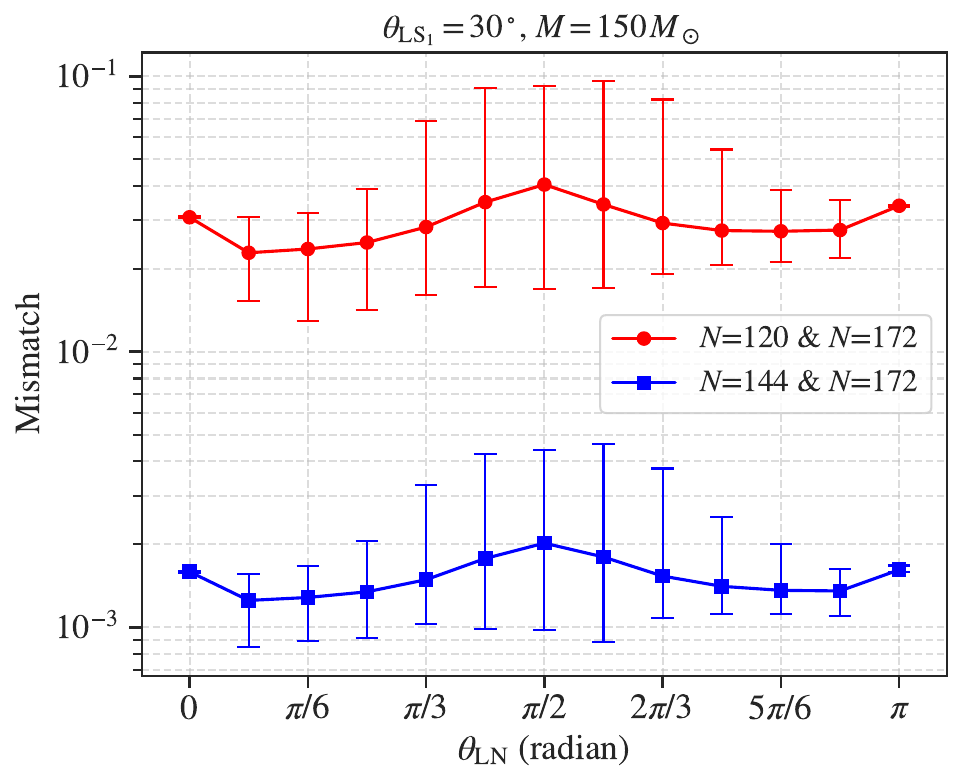}
    \caption{Mismatch between NR waveforms for {\tt CF\_{81}} configuration ($\{q,\, \chi_1,\,  \theta_{\mathrm{LS}_{1}} \}=\{18, 0.8, 30^\circ \}$) at different numerical resolutions as a function of the binary inclination angle $\theta_{\rm LN}$ (angle between the line of sight and the initial orbital angular momentum) at redshifted total mass $150 M_{\odot}$. The mismatch is optimized over luminosity distance, coalescence time and phase, sky location, and polarization. Each solid marker denotes the SNR-weighted mismatch, while the vertical lines show the range between the minimum and maximum mismatch obtained across the sampled signal polarizations and phases (see text for details).
    }
    \label{fig:NRNR_Mismatch_Theta_LS_30_Mtot_150}
\end{figure}

\subsection{Accuracy of NR simulations}\label{sec:nraccuracy} 
To evaluate the accuracy of the NR simulation data at such a large mass ratio, we study the configuration ${\tt CF\_{81}}$, with parameters $\{q,\, \chi_1,\,  \theta_{\mathrm{LS}_{1}} \}=\{18, 0.8, 30^\circ \}$. The two primary sources of error in our NR waveforms are the finite numerical resolution of the simulation and the finite radius at which the waveform data are extracted. To evaluate the impact of finite resolution, we performed three simulations of this configuration at low, medium, and high resolutions. These resolutions correspond to $N=\{120, 144, 172\}$ grid points within the refinement boxes that surround the punctures. The width of the smallest refinement box around each BH is typically of the order of $\sim 2m/N$, where $m$ is the mass of the BH. The waveform data are extracted at a distance of $90M$ from the source for all three resolutions. To quantify errors in the NR waveforms arising from the finite resolution of the simulation, we estimate the mismatch between the highest-resolution ($N=172$) waveform and those at medium ($N=144$) and lower ($N=120$) resolutions.

In the mismatch calculation, we adopt a redshifted total mass of $M=150M_{\odot}$ and estimate the optimized mismatch as outlined in Sec.~\ref{sec:mismatch}. Note that here we cannot optimize the mismatch over rotations of the in-plane spin components of the template. 
Therefore, the match is optimized over all extrinsic parameters, excluding the binary inclination angle $\theta_{\rm LN}$, while the intrinsic parameters (masses and spin parameters) are kept fixed according to the NR simulation configuration.
We present the mismatches as a function of the inclination angle $\theta_{\rm LN}$ in Fig.~\ref{fig:NRNR_Mismatch_Theta_LS_30_Mtot_150}. Solid markers indicate the SNR-weighted mismatch, while vertical lines denote the range of mismatches across the sampled polarizations and phases. For example, for $\theta_{\rm LN}=\pi/3$, the smallest mismatch between the $N=120$ and $N=172$ NR waveforms is 0.016, while the smallest
mismatch between the $N=144$ and $N=172$ waveforms is 0.001. The average mismatch is 0.03 between the $N=120$ and $N=172$ simulations, and 0.002 between the $N=144$ and $N=172$ simulations. We note that these are much larger than the estimate of the mismatch error for the simulations in Ref.~\cite{Hamilton:2023qkv}, which was $\sim$$10^{-4}$ (see Fig.~8 of Ref.~\cite{Hamilton:2023qkv}) for a binary with total mass $100M_\odot$.

Adopting the Richardson extrapolation scheme, Ref.~\cite{Hamilton:2023qkv} derived a relation [their Eq.~(20)] describing how the mismatch between two NR waveforms is expected to scale with an expansion parameter (e.g., numerical resolution or waveform extraction radius) characterizing the NR waveforms, assuming a given convergence order (see also Ref.~\cite{Ferguson:2020xnm}). 
If we perform simulations with two different numerical resolutions characterized by expansion parameters $\Delta_1 = 1/N_1$ and $\Delta_2 = 1/N_2$ (where the moving boxes around the punctures have $N^3$ points) and assume a convergence order $n$, then the mismatch between the waveforms $h_1$ and $h_2$, computed from the two simulations, is given by
\begin{equation}
\mathfrak{M}(h_1,h_2) = \kappa \left(\Delta_1^n - \Delta_2^n\right)^2,
\end{equation}
where $\kappa$ is a constant.
From the two average mismatch values quoted above from Fig.~\ref{fig:NRNR_Mismatch_Theta_LS_30_Mtot_150}, we find a consistent value of $\kappa$ for $n\approx 6$, which implies sixth-order convergence for our simulations. 
The code uses sixth-order accurate spatial finite differencing, but fourth-order accurate Runge-Kutta time-stepping. At sufficiently high resolution, we expect the fourth-order time-stepping error to dominate.
Observing apparent sixth-order convergence may indicate that the time-stepping error does not yet dominate, or—more likely given the simulation length—that the system is not yet in the fully convergent regime.

Assuming sixth-order convergence, the mismatch between the waveform from the $N=172$ simulation and the true waveform at infinite resolution ($N \rightarrow \infty$) is estimated to be $\sim 6\times10^{-4}$, while for the waveform from the $N=120$ simulation it is $\sim 0.04$.
This suggests that the $N=120$ simulation is not suitable for high-precision studies.
Nevertheless, we will see later that, despite its lower resolution, the $N=120$ simulation captures the broad physical features of the waveform reasonably well.
The $N=172$ simulation, by contrast, is likely of comparable accuracy to simulations in the previous BAM catalogue~\cite{Hamilton:2023qkv}.

\begin{figure}[t]
    \centering
    \includegraphics[width=0.975\columnwidth]{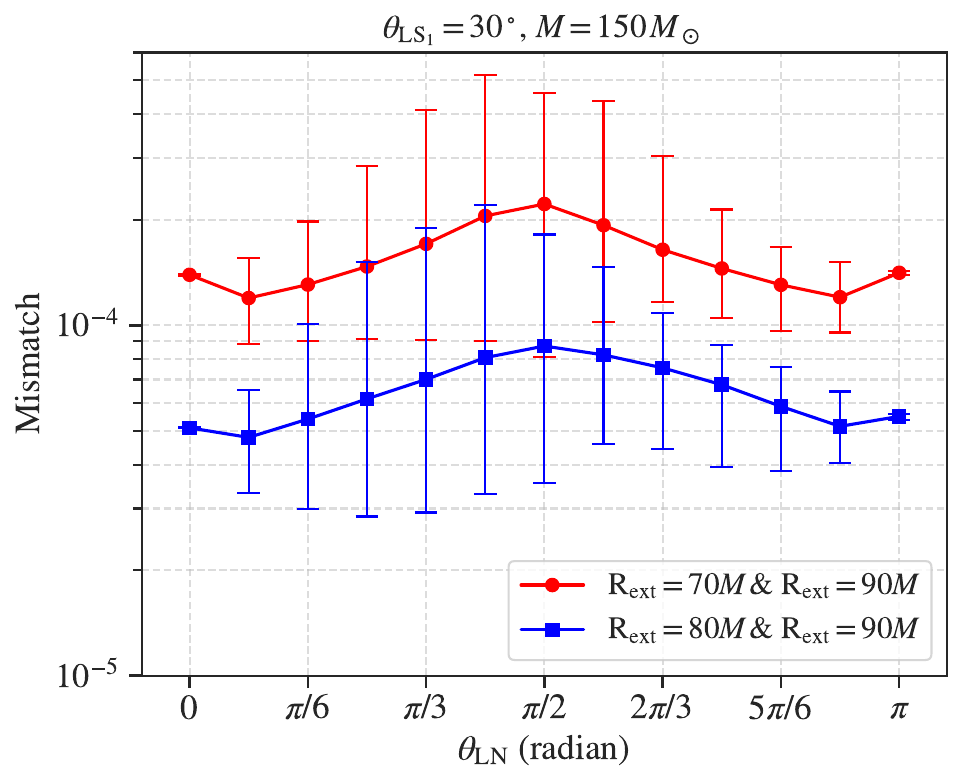}
    \caption{Mismatch between NR waveforms for ${\tt CF\_{81}}$ configuration in Table~\ref{table:metadata} at different extraction radii (as indicated in the legend) as a function of the binary inclination angle $\theta_{\rm LN}$ at redshifted total masses of $150 M_{\odot}$.
    The parameters over which the mismatch is optimized, as well as the plotting style and interpretation, are the same as in Fig.~\ref{fig:NRNR_Mismatch_Theta_LS_30_Mtot_150}.
    }
    \label{fig:NRNR_Rext_Mismatch_Theta_LS_30_Mtot_150}
\end{figure}

To further evaluate the impact of waveform extraction at finite radius, we consider the $N=120$ simulation and extract the waveform at three different radii from the source: ${\rm R_{ext}} = 70M$, $80M$, and $90M$. To quantify errors in the NR waveforms arising from finite-radius extraction, we perform a mismatch study similar to the previous analysis of finite-resolution errors, but now comparing the waveform extracted at ${\rm R_{ext}} = 90M$ with those extracted at ${\rm R_{ext}} = 70M$ and $80M$. The results are shown in Fig.~\ref{fig:NRNR_Rext_Mismatch_Theta_LS_30_Mtot_150}. 

As anticipated, the mismatches between the waveforms extracted at ${\rm R_{ext}} = 80M$ and ${\rm R_{ext}} = 90M$ are smaller than those between ${\rm R_{ext}} = 70M$ and ${\rm R_{ext}} = 90M$, typically by a factor of $\sim2–5$. The average mismatch is approximately $2\times10^{-4}$ for the $70M$ vs.\ $90M$ mismatches, and $7\times10^{-5}$ for the $80M$ vs.\ $90M$ mismatches. For comparison, these values are again notably larger than the estimated mismatch error reported in Ref.~\cite{Hamilton:2023qkv}, which was $\sim 10^{-5}$ (see Fig.~9 of Ref.~\cite{Hamilton:2023qkv}) for a binary with total mass $100M_\odot$.

From this mismatch study, we infer first-order convergence for the finite-radius extraction error, i.e., the error due to finite extraction radius falls of as $\sim1/R_{\rm ext}$. Assuming first-order convergence, we estimate the mismatch between the waveform extracted at $\rm R_{ext} = 90M$ and the true waveform at ${\rm R_{ext}} \rightarrow \infty$ to be $\sim 4\times10^{-3}$. Using Eq.~(23) of Ref.~\cite{Hamilton:2023qkv}, we estimate the mismatch between our NR waveforms—computed at finite resolution ($N=120$) and extraction radius (${\rm R_{ext}}=90M$)—and the true waveform at infinite resolution and extraction radius to be $(\sqrt{\mathfrak{M}_{\rm resolution}}+\sqrt{\mathfrak{M}_{\rm extraction}})^{2}\approx(\sqrt{0.04}+\sqrt{0.004})^{2}\approx0.07$. This corresponds to an upper bound of $\sim7\%$ on the mismatch error of the reported NR waveforms. The overall mismatch error for the higher-resolution simulations is 0.016 for $N=144$ and 0.007 for $N=172$. (For the NR simulations in Ref.~\cite{Hamilton:2023qkv}, the overall mismatch error was estimated at $\sim 4\times10^{-3}$ for a binary with total mass $100M_\odot$.)

Nominally, these mismatch errors are large. The conservative indistinguishability criterion ($\mathfrak{M}\le1.35/\rho^{2}$)~\cite{Baird:2012cu} suggests that the $N=120$ waveform could be distinguished from a true waveform at any SNR above 5, i.e., any detectable signal. (For the $N=172$ waveform, it is any SNR above 14.) However, as illustrated in Ref.~\cite{Thompson:2025hhc}, this criterion is generally too conservative, and the actual indistinguishability SNR typically occurs at much higher values. We will see in Sec.~\ref{sec:accuracy-pe} that this is also the case for these waveforms; in fact, the two highest resolutions are almost completely indistinguishable at an SNR of 50, and would likely not be distinguishable even at an SNR of 100.

To achieve comparable overall mismatch error to the $q\leq 8$ precessing-binary simulations in the earlier BAM catalog~\cite{Hamilton:2023qkv} (0.004), we would require each of our simulations to be performed with higher resolution than our $N=172$ simulation (0.007). With our current methods the computational cost would be prohibitive. The $N=120$ simulation required 1.6M core hours, while the $N=172$ simulation, by far the most expensive we have performed to date, required almost an order of magnitude more computational resources, 13.2M CPU hours. We will discuss the implications of this further in Sec.~\ref{sec:conclu}.

\begin{figure}[t]
    \centering
\includegraphics[width=0.48\textwidth]{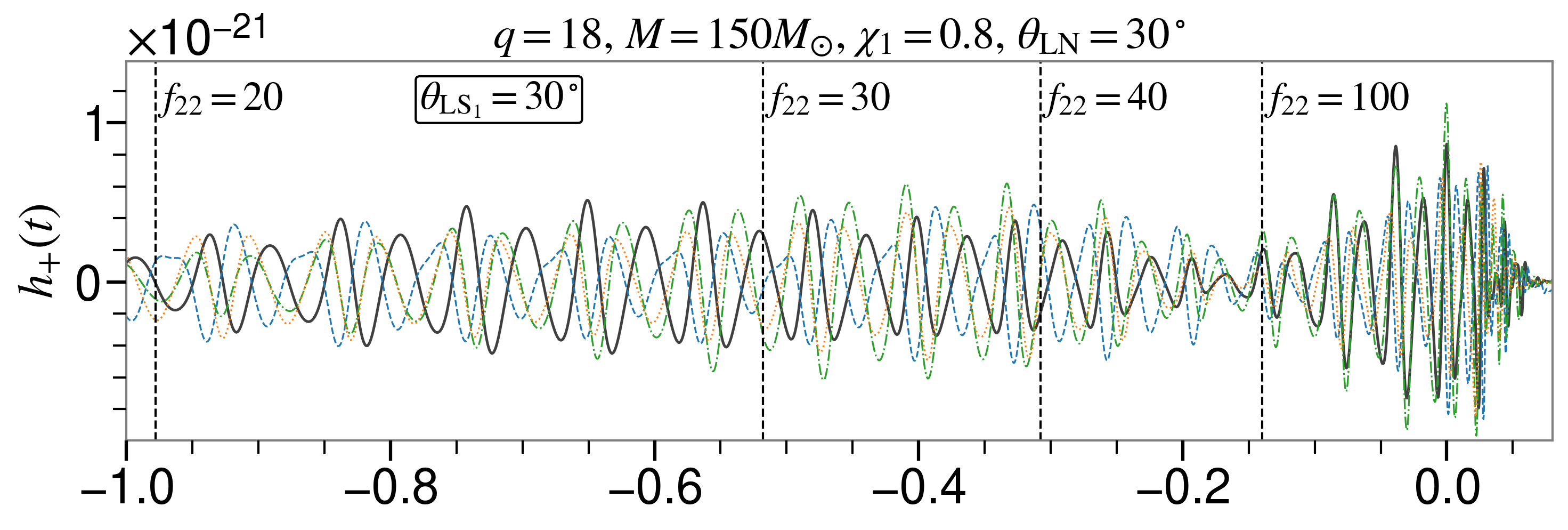}\\
\includegraphics[width=0.485\textwidth]{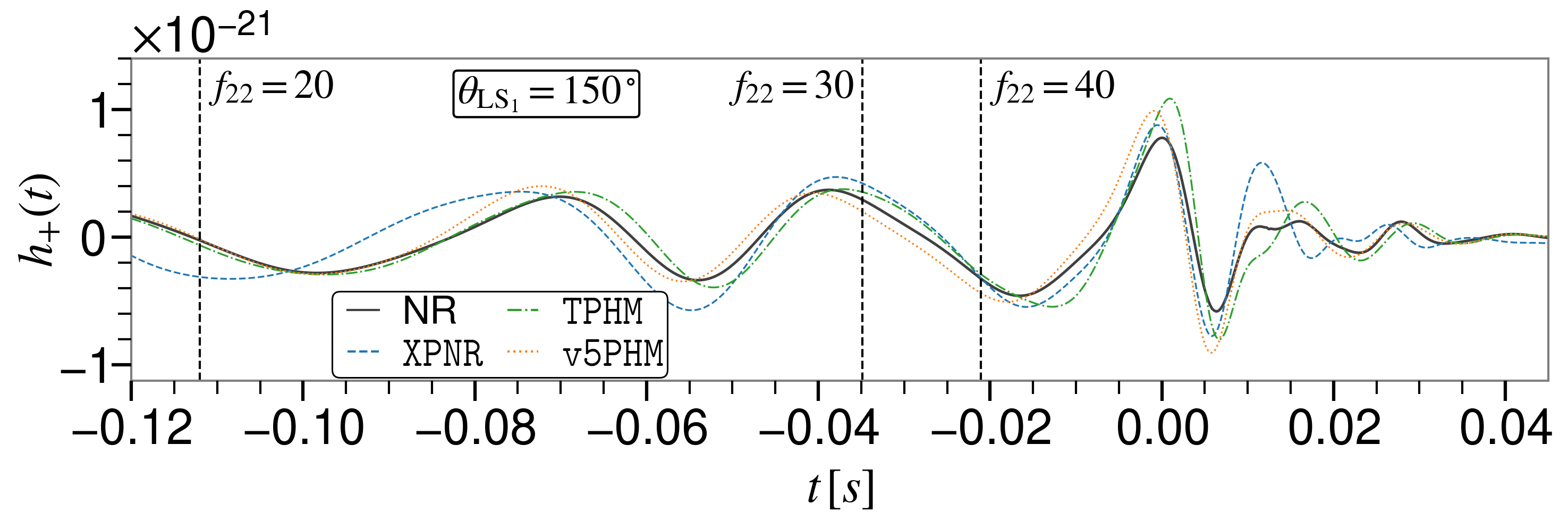}
\caption{Plus polarization of NR simulations and three contemporary waveform models, {\tt XPNR}, {\tt TPHM}, and {\tt v5PHM}. Redshifted total mass is $150 M_{\odot}$, $\theta_{\mathrm{LN}}=30^\circ$, and $D_L=300$ Mpc (with the polarization angle chosen to be zero).
Top: NR simulation ${\tt CF\_{81}}$ ($\theta_{\rm LS_1} = 30^\circ$). Bottom: ${\tt CF\_{85}}$ ($\theta_{\rm LS_1} = 150^\circ$).
The {\tt XPNR} (blue dashed), {\tt TPHM} (green dash-dotted), and {\tt v5PHM} (orange dotted ) models show significant disagreement with the underlying NR data (solid black).
}
    \label{fig:hp_cmprsn}
\end{figure}
\section{Mismatches against current models}\label{sec:accuracy-mismatch}

We now quantify the accuracy of current waveform models with respect to our new NR simulations. 
Specifically, we assess the accuracy of three contemporary waveform models ({\tt XPNR}, {\tt TPHM}, and {\tt v5PHM}) by computing the mismatches following the procedure outlined in Sec.~\ref{sec:mismatch}. We again consider the redshifted total mass of $M=150M_{\odot}$ for the mismtach calculation. 
We choose this mass to ensure that the detector's sensitivity band contains a full signal for all $\ell\leq4$ modes. 

In Fig.~\ref{fig:hp_cmprsn}, we show the plus polarizations for the NR simulations ${\tt CF\_{81}}$ and ${\tt CF\_{85}}$ (which are used in the parameter estimation studies in Sec.~\ref{sec:accuracy-pe}), choosing a redshifted total mass of $150 M_{\odot}$ and $\theta_{\mathrm{LN}}=30^\circ$, with the polarization angle set to zero and the luminosity distance fixed at 300 Mpc. We superimpose the NR waveform with predictions from the three waveform models. 
We optimize over the the reference time, reference phase and in-plane spin components to identify the best-matching template. All models look significantly different compared to the NR waveform.

\begin{figure}[hb]
    \centering
    \includegraphics[width=0.975\columnwidth]{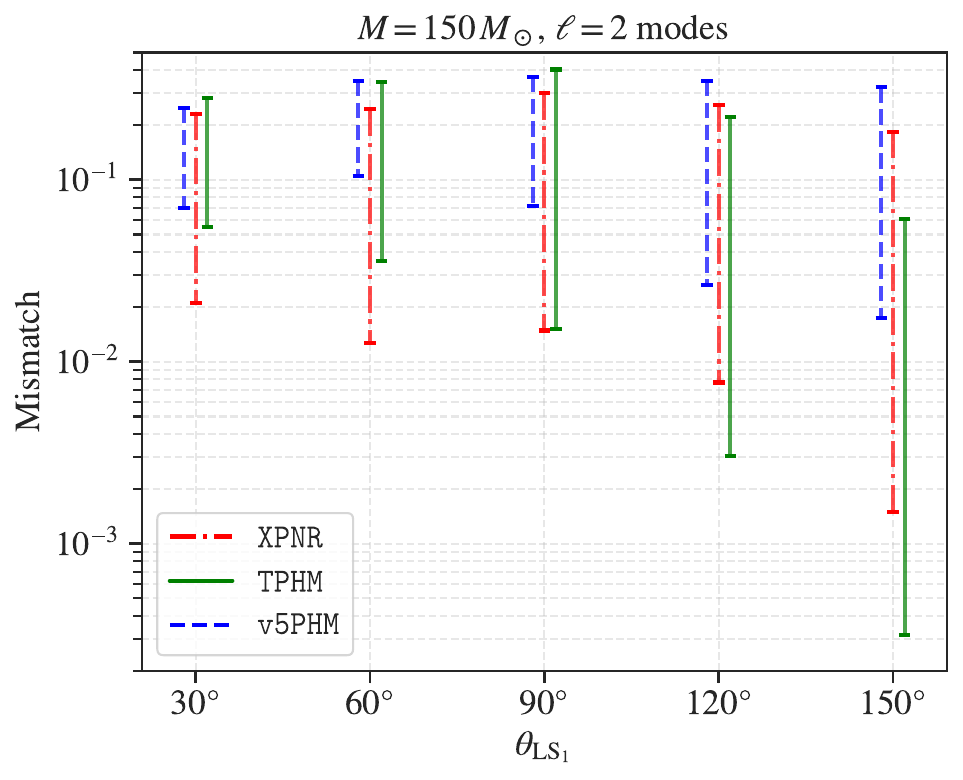}\\
    \includegraphics[width=0.975\columnwidth]{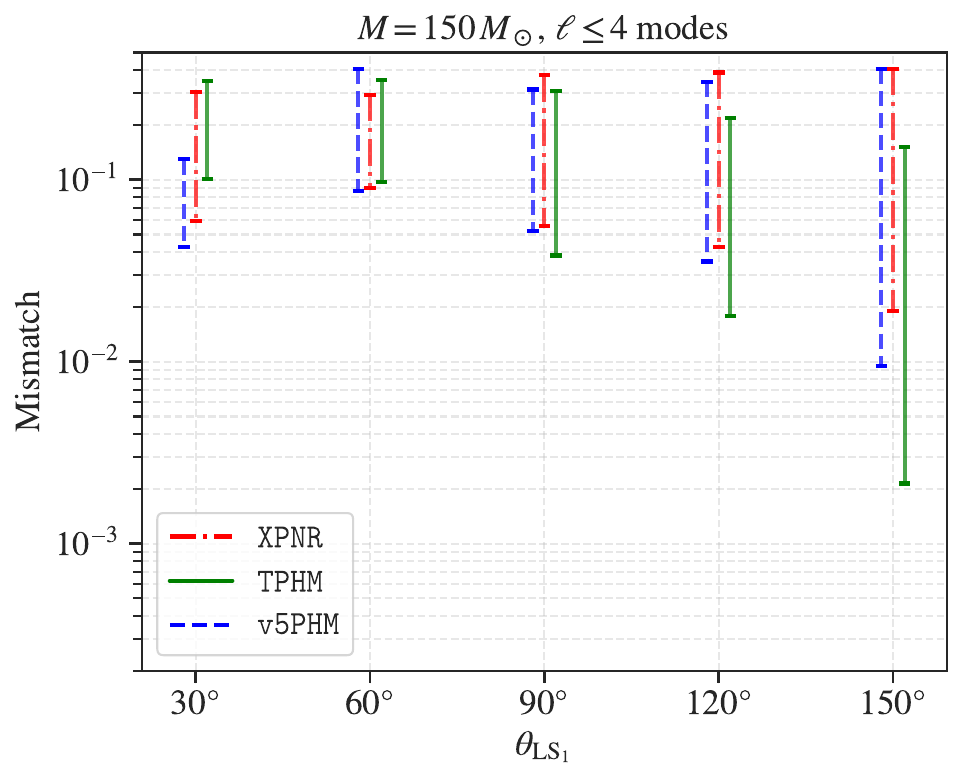}
    \caption{Mismatches against three current models. 
    The spin misalignment angle $\theta_{\rm LS_{1}}$ of each simulation is indicated on the $x$-axis. The range of values corresponds to variations in inclination, polarisation and reference phase. Top: mismatches restricted to $\ell =2$ modes of both the NR waveforms and the waveform models. Bottom: all available modes in both NR and model.
    }
    \label{fig:mismatches-summary}
\end{figure}

We then quantify these differences using the mismatch metric described in Sec.~\ref{sec:mismatch}, providing a more systematic measure of the waveform models' accuracies across the parameter space covered by our NR simulations.
The mismatches are optimized over luminosity distance, coalescence time, and phase, sky location, polarization, and precession angle.
The optimized mismatch is computed across a grid of signal inclination, polarization, and reference phase values, with each parameter varied over its full physical range as described in Sec.~\ref{sec:mismatch}. For each NR–model comparison, this procedure yields a collection of optimized mismatch values spanning all sampled parameter combinations. In our results we show the full range of these values. 

The mismatches are shown in Fig.~\ref{fig:mismatches-summary}. In the top panel we restrict to the $\ell = 2$ modes in both the waveform models and the NR simulations. In the bottom panel we include all available modes in the waveform models and modes up to $\ell=4$ in the NR simulations. We have not considered the memory modes (modes with $m=0$) from the NR simulations; the available modes in each model are listed in Sec.~\ref{sec:intro}.

We find that the mismatches are poor for all configurations. For all models there are configurations where the mismatch is as high as 0.4, i.e., a match of 0.6; these are extremely poor matches that, in the comparable-mass regime, would disqualify a model from use in parameter estimation. High mismatches are found even when restricting to only the $\ell=2$ modes. Overall we see some improvement for larger values of $\theta_{\rm LS_1}$, but recall that these configurations have large anti-aligned spin components, leading to merger at lower frequencies (see Fig.~\ref{fig:NR_sim}), and therefore far fewer GW cycles from the mismatch starting frequency $f = 20\,{\rm Hz}$. Note also that these mismatches are typically much larger than the mismatch errors of the NR waveforms. For the $\theta_{\rm LS_1} = 30^\circ$ configuration, the one configuration where we have full NR error estimates (Sec.~\ref{sec:nraccuracy}), the mismatch error is 0.007 for the highest-resolution simulation, well below the smallest mismatch (0.02) against any model. We have also found that for this configuration the range of mismatches against the models do not vary significantly with NR resolution. 

We also show the optimized mismatch as a function of the binary orbital inclination angle $\theta_{\rm LN}$, for two redshifted total masses in the Supplemental Materials.

The mismatches suggest that none of the current models is able to accurately capture the full complexity of the GW signals in our NR simulations, even when considering the dominant quadrupolar modes. The discrepancies become more pronounced when higher-order modes are included, suggesting that the treatment of these modes in the waveform models remains inadequate, although it is by no means the only source of error. (Note that for aligned-spin systems, the worst mismatches against NR waveforms in Fig.~16 of Ref.~\cite{Garcia-Quiros:2020qpx}, which includes configurations at $q=18$, is $\sim$0.01.)
We also see that the NR-model mismatches are typically much larger than the mismatch uncertainties in the NR waveforms, indicating that the large mismatches cannot be ascribed to NR errors. 
We draw the clear conclusion that significant improvements are required in the modeling of high-mass-ratio precessing binaries.

\begin{figure*}[htb!]
    \centering
    \includegraphics[width=0.4955\textwidth]{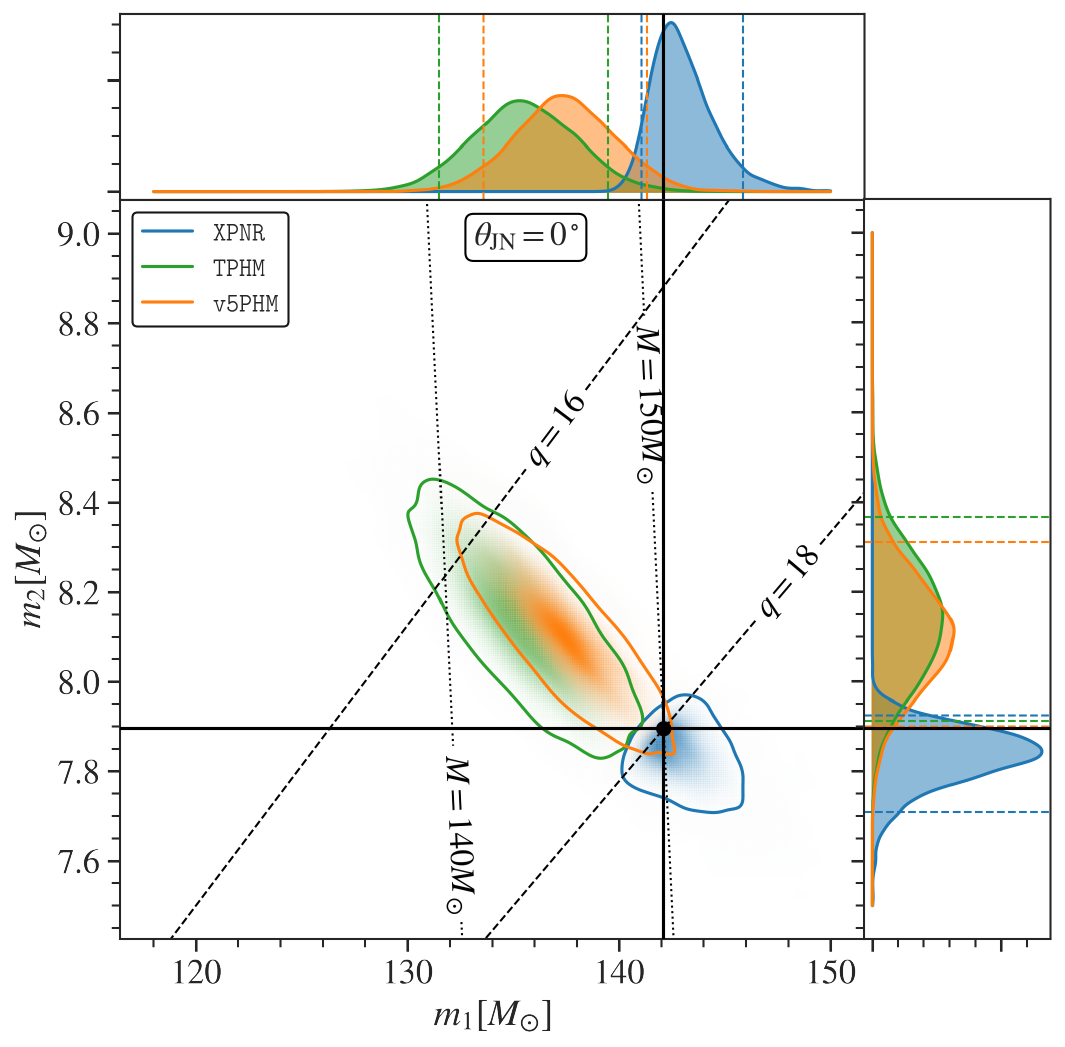}
    \includegraphics[width=0.4955\textwidth]{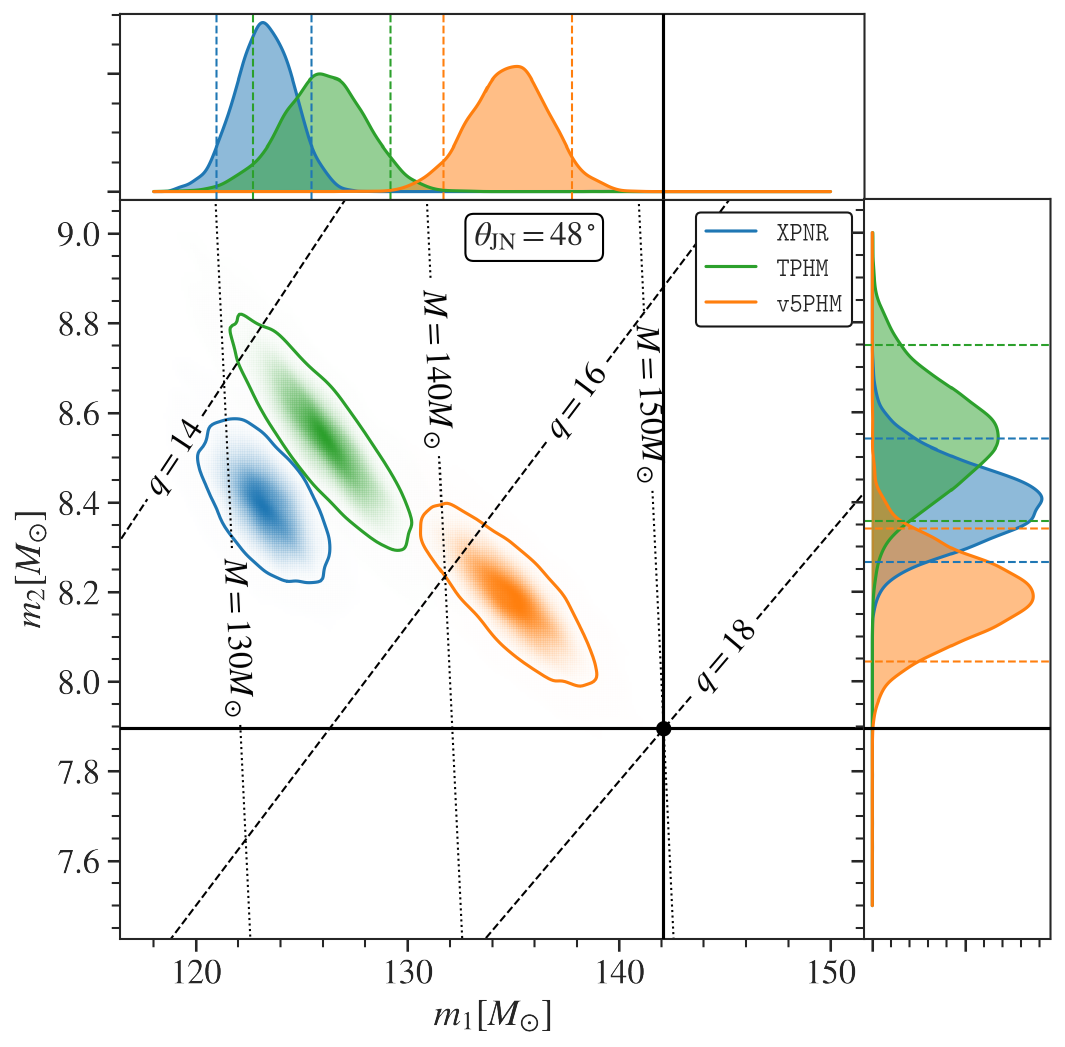}\\
    \includegraphics[width=0.4955\textwidth]{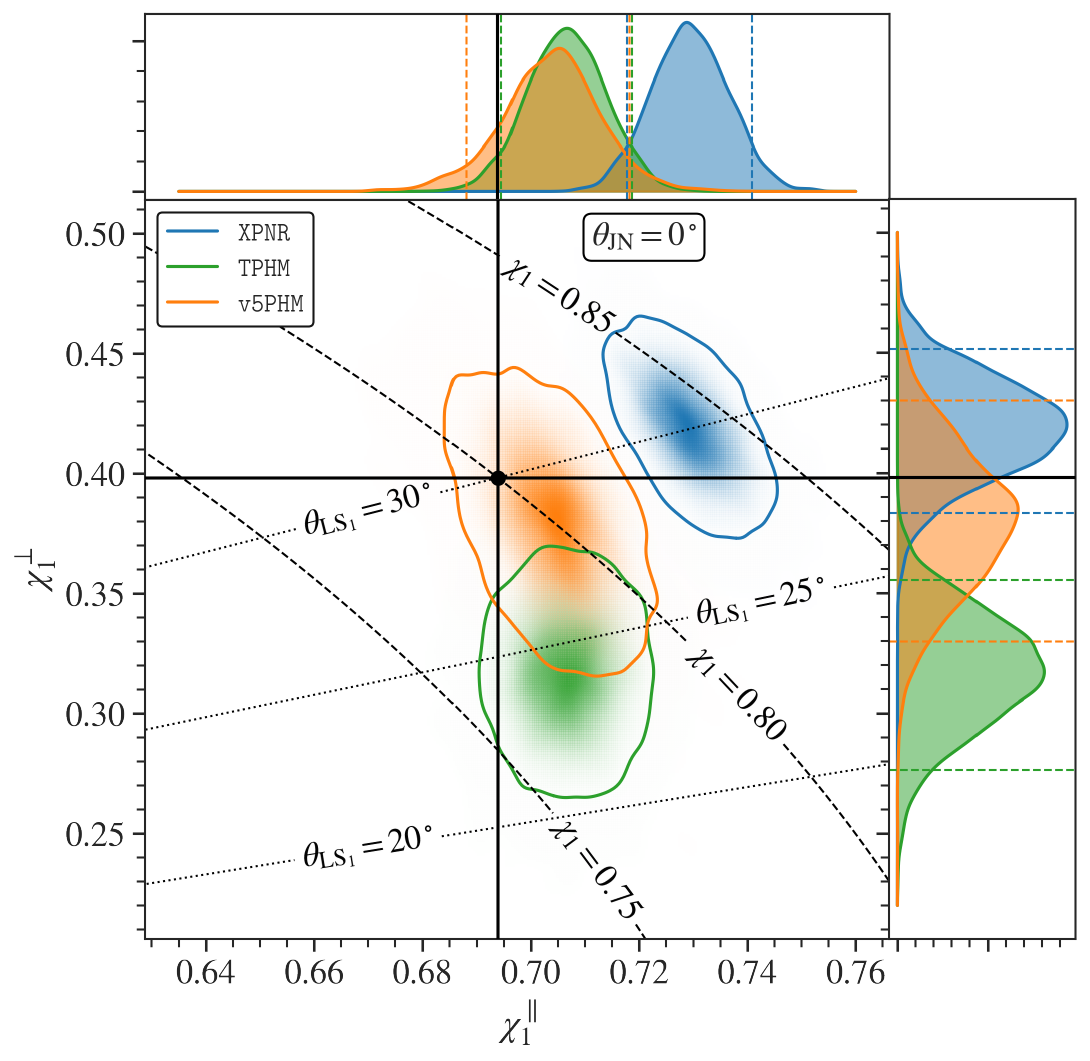}
    \includegraphics[width=0.4955\textwidth]{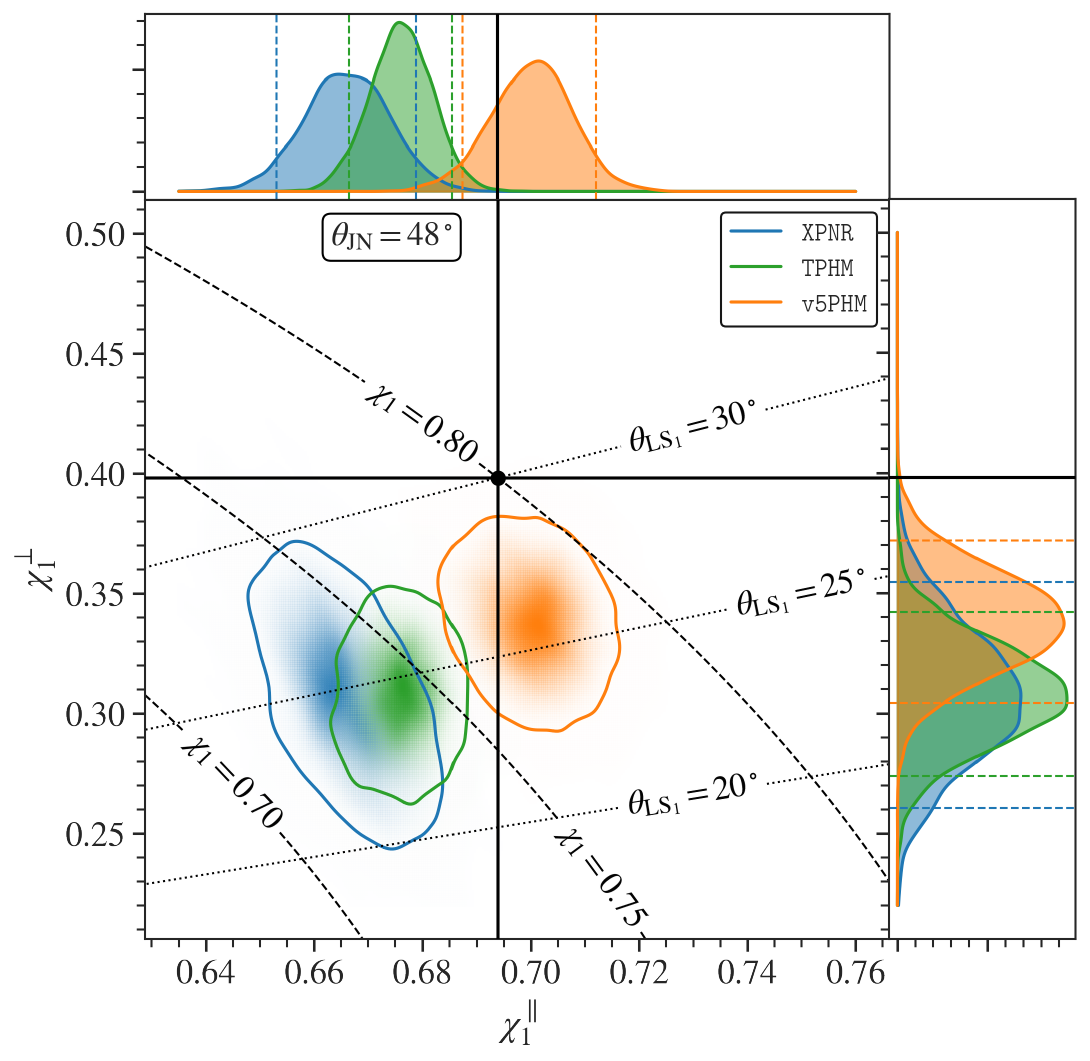}
    \caption{Mass and primary spin measurements of two signals generated from the $\theta_{\text{LS}_1}=30^\circ$ simulation (${\tt CF\_{81}}$ in Table~\ref{table:metadata}) with a redshifted total mass of $150M_{\odot}$ and SNR 50, using the {\tt XPNR}, {\tt TPHM}, and {\tt v5PHM} waveform models. 
    Left: $\theta_{\text{JN}} = 0^\circ$. Right: $\theta_{\text{JN}} = 48^\circ$. 
    Top panels: component masses ($m_1$ and $m_2$). Bottom panels: primary spin components along the orbital angular momentum ($\chi_{1}^{\parallel}\equiv\chi_{1} \cos \theta_{\rm LS_{1}}$) and in the orbital plane ($\chi_{1}^{\perp}\equiv \chi_{1} \sin \theta_{\rm LS_{1}}$), at the reference frequency.
    The black dot and lines indicate the true (injected) parameters. The contours and dashed lines represent the 90\% credible regions. The black dashed and dotted curves denote contours of constant mass ratio and redshifted total mass (top), and primary spin magnitude and $\theta_{\text{LS}_1}$ (bottom).
    }
    \label{fig:Theta_LS_30_Theta_JN_0_48}
\end{figure*}

\begin{figure*}[t]
    \centering
    \includegraphics[width=0.4955\textwidth]{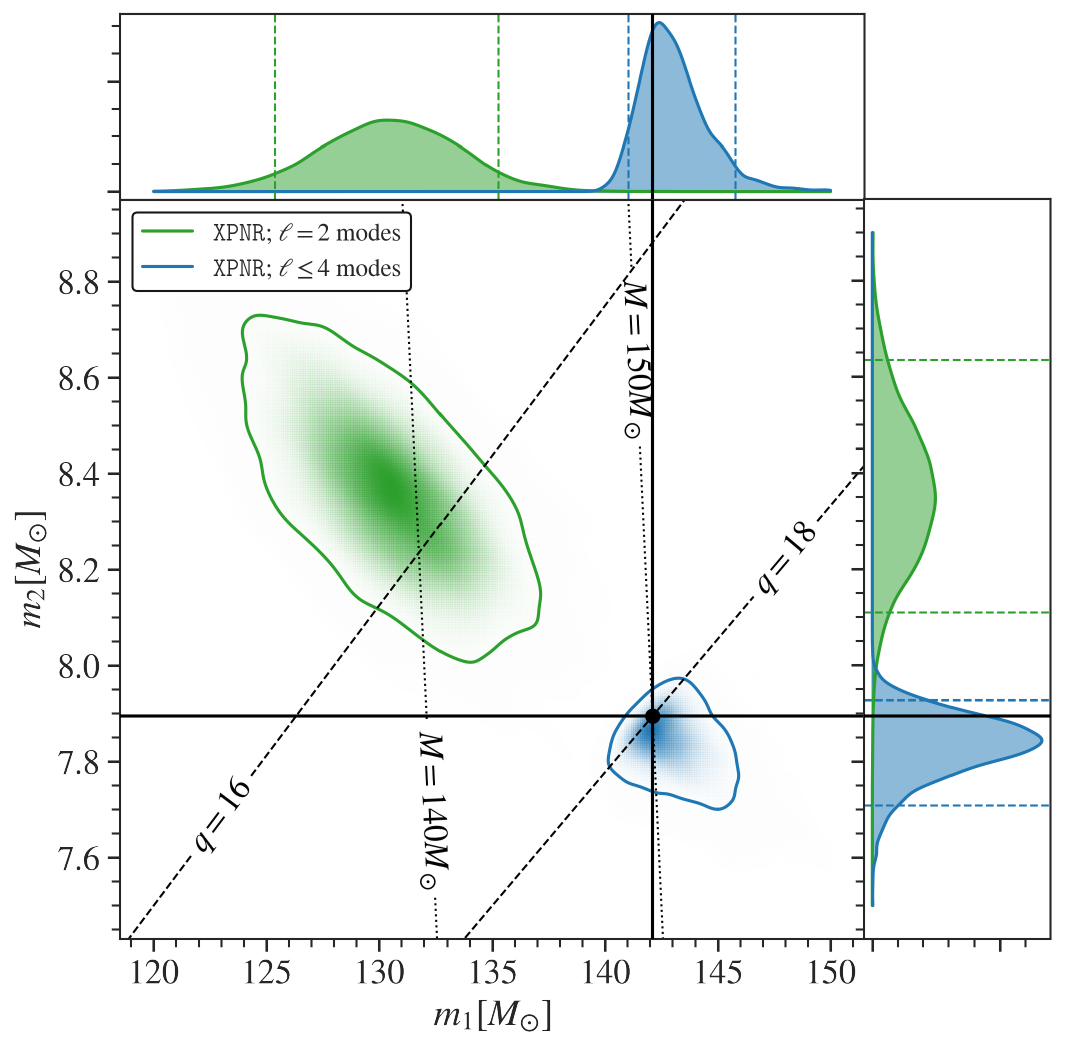}
    \includegraphics[width=0.4955\textwidth]{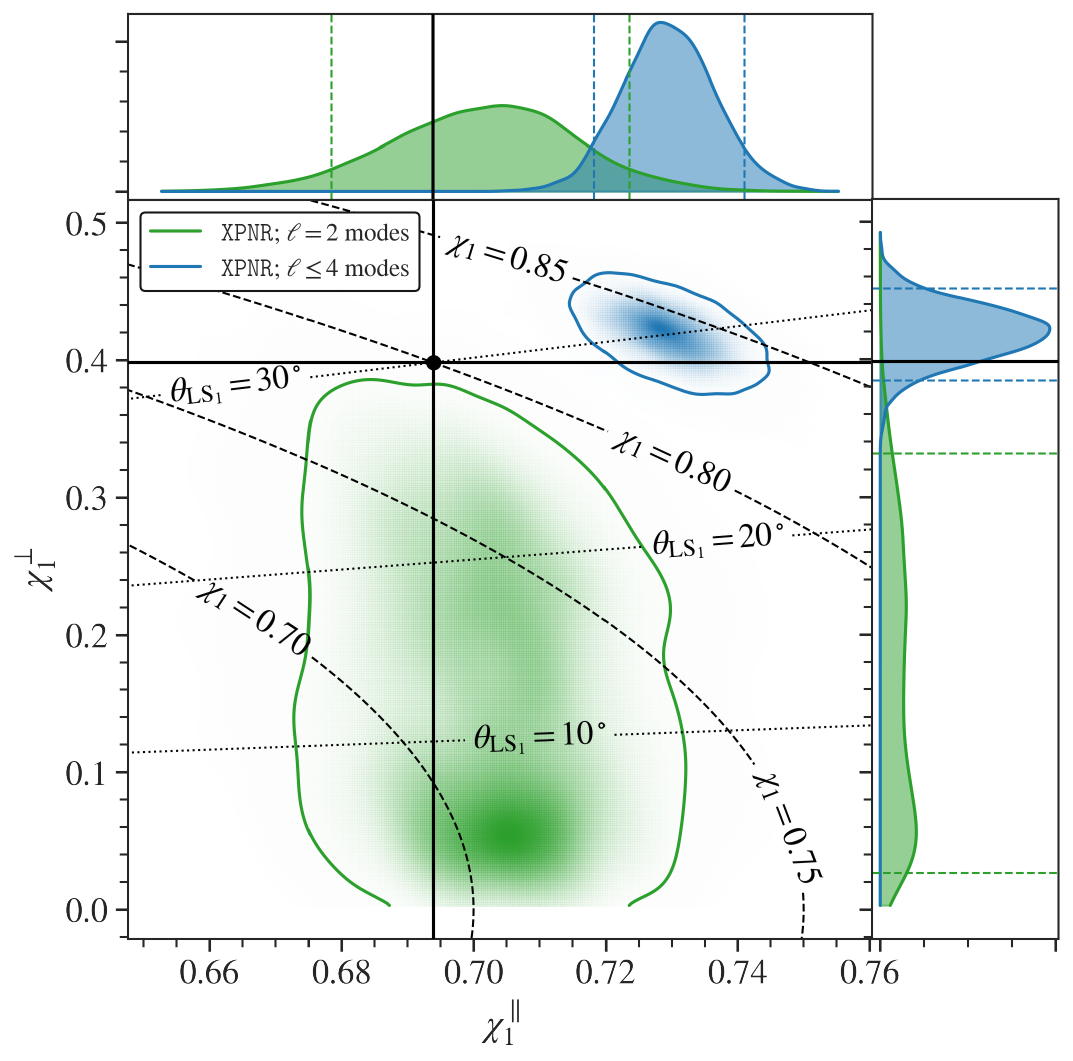}
    \caption{Mass and primary spin measurements for the same $\theta_{\rm LS_1} = 30^\circ$, $\theta_{\rm JN} = 0^\circ$ signal as in Fig.~\ref{fig:Theta_LS_30_Theta_JN_0_48}. In the blue results, only the quadrupolar modes (i.e., $\ell = 2$ modes) are used for both the NR injection and the parameter estimation with the {\tt XPNR} model. In the green results, the full NR signal is again recovered with the full {\tt XPNR} model.}
    \label{fig:Theta_LS_30_Theta_JN_0_XPNR_mode_cpmrsn}
\end{figure*}

\begin{figure*}[t]
    \centering
    \includegraphics[width=0.4955\textwidth]{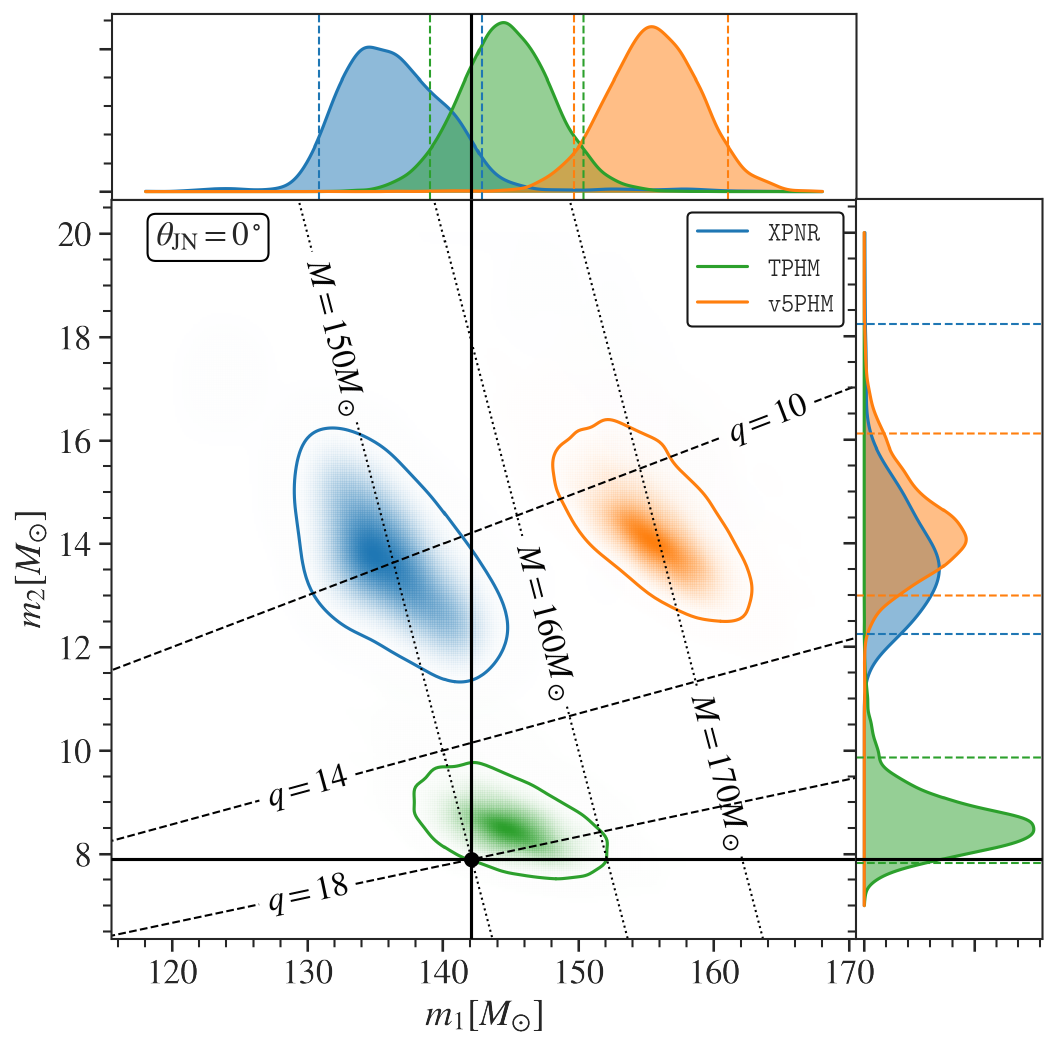}
    \includegraphics[width=0.4955\textwidth]{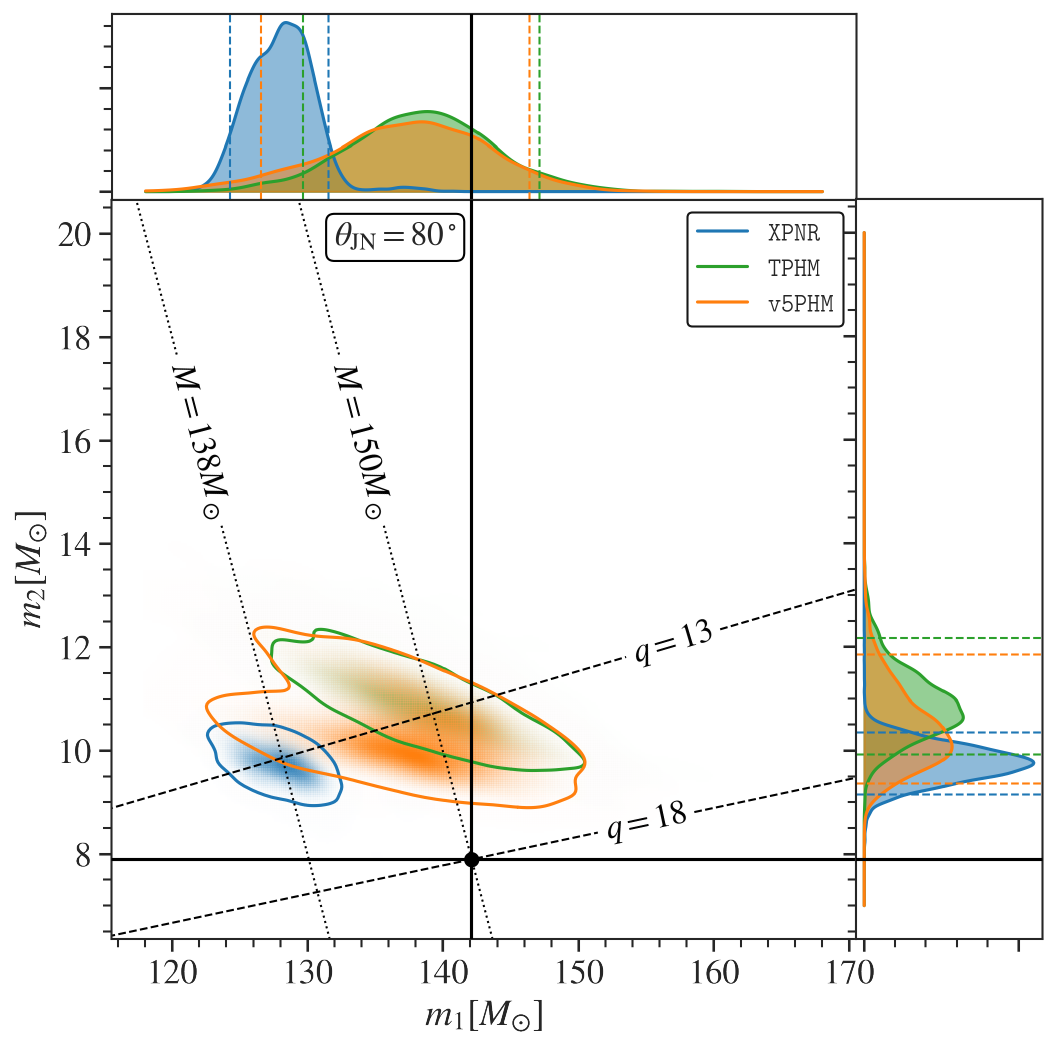}\\
    \includegraphics[width=0.4955\textwidth]{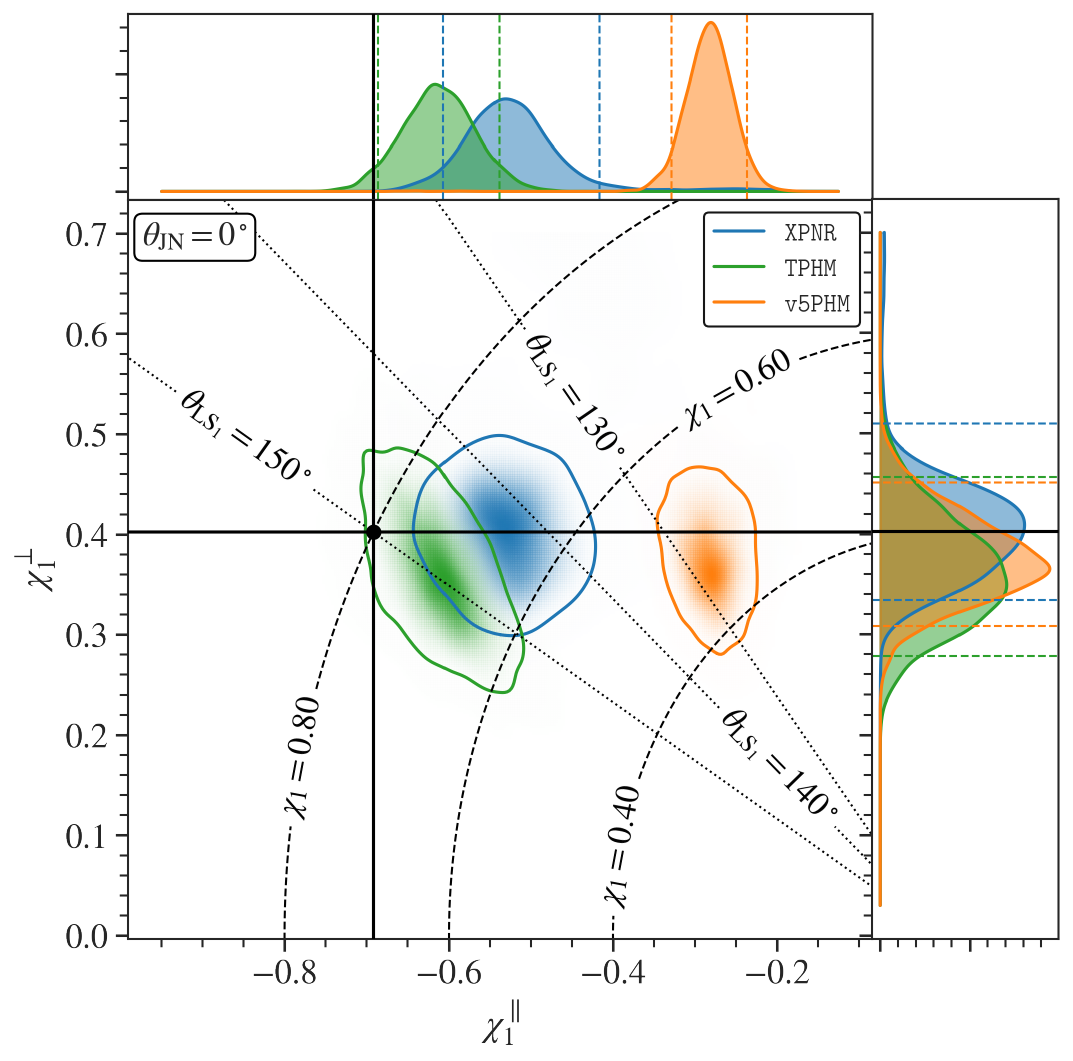}
    \includegraphics[width=0.4955\textwidth]{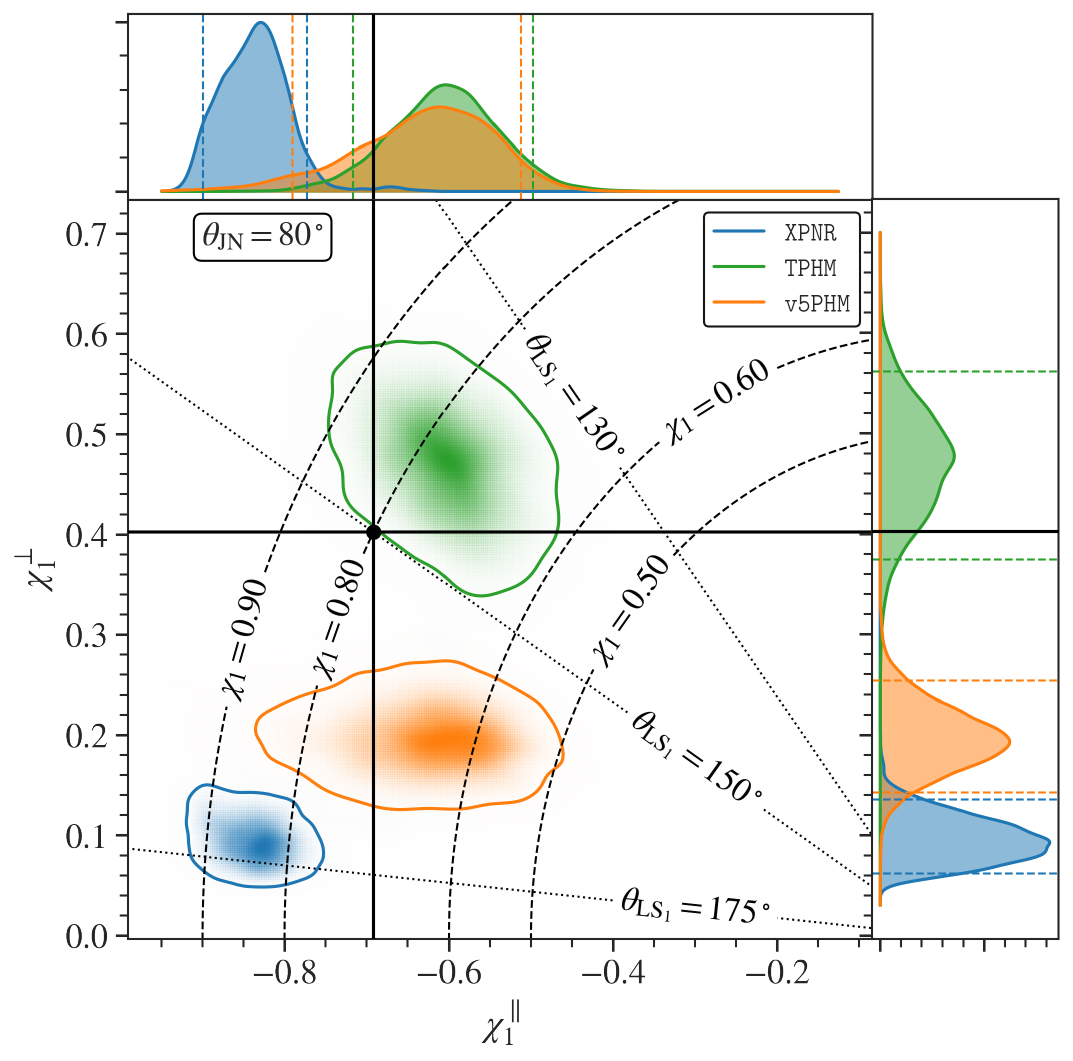}
    \caption{    
    Mass and primary spin measurements for the ${\tt CF\_85}$ ($\theta_{\rm LS_1}=150^\circ$) configuration with redshifted total mass $150M_{\odot}$ and SNR 50, for orientations $\theta_{\rm JN}=0^\circ$ (left) and $\theta_{\rm JN}=80^\circ$ (right).
    }
    \label{fig:Theta_LS_150_Theta_JN_0_80}
\end{figure*}

\begin{figure*}[ht]
    \centering
    \includegraphics[width=0.4955\textwidth]{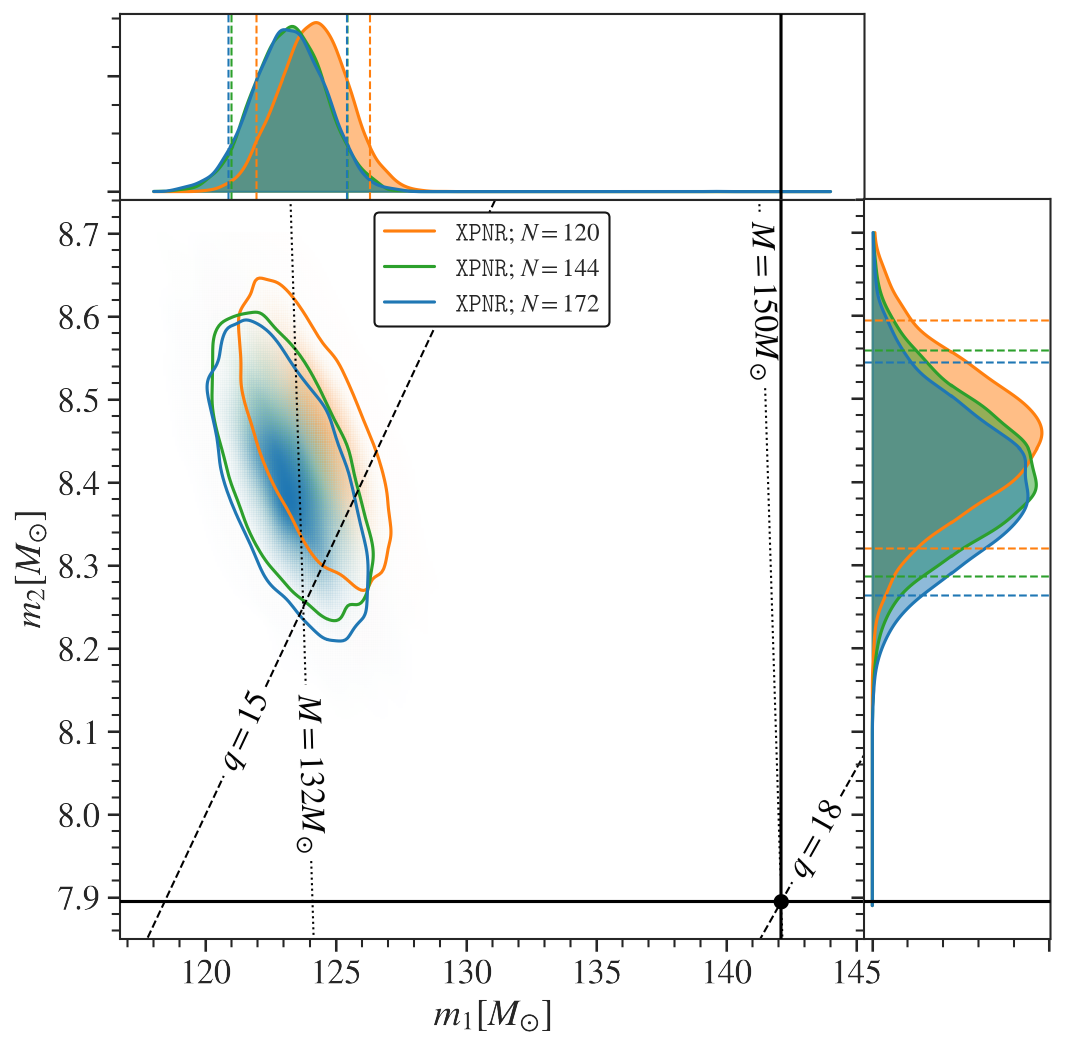}
    \includegraphics[width=0.4955\textwidth]{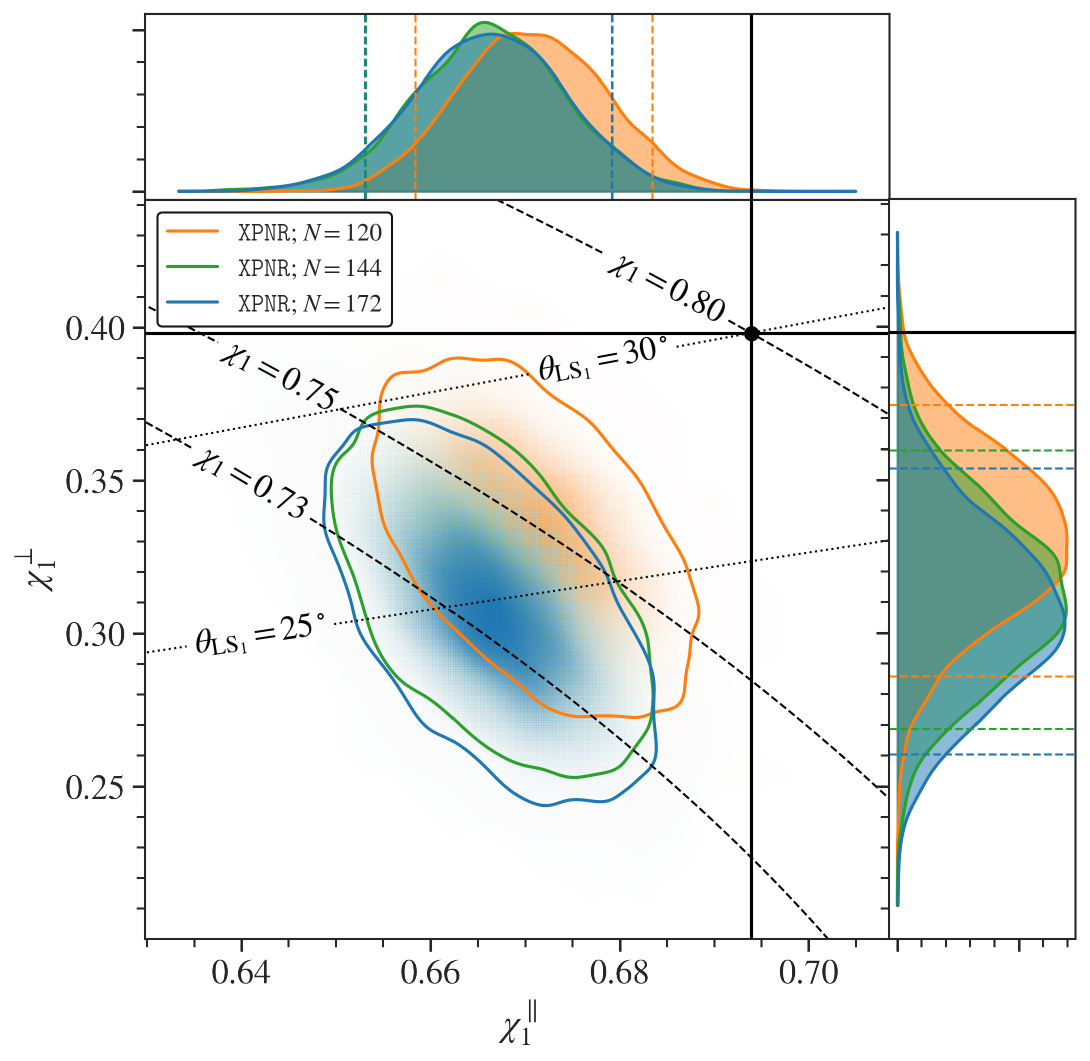}
    \caption{
    Mass and primary spin measurements for the $\theta_{\rm LS_1} = 30^\circ$ system with redshifted total mass $150M_{\odot}$ and orientation $\theta_{\text{JN}}=48^\circ$, from the NR simulations at three resolutions, $N=120$, $N=144$, and $N=172$, recovered with the {\tt XPNR} model.
    }
    \label{fig:Theta_LS_30_Mtot_150_Theta_JN_48_XPNR_N_120_144_172}
\end{figure*}

\section{Parameter estimation with current models}\label{sec:accuracy-pe}

Mismatches are a standard metric to assess the accuracy of waveform models, but they do not tell us which parameters (if any) will be biased when the model is used to measure the source properties following an observation. Therefore, to assess the performance of current models in practice, we have also carried out Bayesian PE on synthetic BBH signals~\cite{Veitch:2014wba} generated from our NR simulations, to assess how well the models recover the source properties. The details of the PE analysis and settings are described in Sec.~\ref{sec:pe}.

We first consider the NR simulation ${\tt CF\_{81}}$ listed in Table~\ref{table:metadata}, with $\theta_{\text{LS}_{1}}=30^\circ$, and generate two synthetic GW signals with a redshifted total mass of $150M_{\odot}$, and an SNR of 50. For the first signal, we set the injected value $\theta_{\text{JN}}=0^\circ$, by adjusting $\theta_{\text{LN}}$ and the reference phase accordingly. For the second signal, we set $\theta_{\text{JN}}=48^\circ$ and choose the reference phase such that $\theta_{\text{LN}}=60^\circ$ at the reference frequency.
Although both configurations have the same intrinsic amount of precession, we expect the former to exhibit a much weaker observable precession imprint on the signal because its line of sight is aligned with the total angular momentum.
We perform PE on both signals using the {\tt XPNR}, {\tt TPHM}, and {\tt v5PHM} waveform models. The posterior distributions of the component masses ($m_1$ and $m_2$) and the primary spin components, both aligned with the orbital angular momentum ($\chi_{1}^{\parallel} \equiv\chi_{1} \cos \theta_{\rm LS_{1}}$) and in the orbital plane ($\chi_{1}^{\perp} \equiv \chi_{1} \sin \theta_{\rm LS_{1}}$)
at the reference frequency, are shown in Fig.~\ref{fig:Theta_LS_30_Theta_JN_0_48}. The left column displays the results for the $\theta_{\mathrm{JN}}=0^\circ$ configuration, while the right column displays those for $\theta_{\mathrm{JN}}=48^\circ$. The top panels show the posteriors on the masses, while the bottom panels show the distributions of the primary spin parameters. We find that the mass and spin parameters are tightly constrained, owing to the presence of precession-induced modulations and the substantial higher-harmonic content~\footnote{For the second signal, the optimal injected SNRs in the $\ell$ = 2, 3, and 4 multipoles are $\sim$ 45, 19, and 8, respectively.}, both of which help break several parameter degeneracies.

From the results in Fig.~\ref{fig:Theta_LS_30_Theta_JN_0_48} we see that no current model reliably infers the correct parameters for the $\theta_{\rm LS_1} = 30^\circ$ system. For the $\theta_{\rm JN} = 0^\circ$ orientation {\tt XPNR} correctly recovers the masses, but the primary spin components are biassed. The true masses are outside the 90\% credible interval for both other models, although {\tt v5PHM} correctly infers the spin components. The biasses at orientation $\theta_{\rm JN} = 48^\circ$ are more severe: no model correctly infers either the masses or the primary spin components. In this case the bias is not small: the errors in the individual-mass recovery are over 10\%, and the recovered mass ratio can be as low as $q=14$.

From the right panel of Fig.~\ref{fig:mismatches-summary}, we see that the minimum possible mismatch between the {\tt XPNR} model and the NR waveform for the $\theta_{\text{LS}_{1}} = 30^\circ$ case is 0.06, which corresponds to a 1D bias SNR of about 5. However, for the $\theta_{\text{JN}} = 0^\circ$ case, we find that even at an SNR of 50, the marginalized 1D posteriors of $m_1$, $m_2$, and $\chi_{1}^{\perp}$ still enclose the true injected values. Even for the $\theta_{\text{JN}} = 48^\circ$ case, a simple SNR rescaling (error bars $\propto 1/\text{SNR}$) suggests that at an SNR of 5, the marginalized 1D posteriors are likely to include the true parameter values. This was also noted in our earlier discussion: the indistinguishability-based SNR estimate is overly conservative, as suggested in Ref.~\cite{Thompson:2025hhc}. We further find that the bias SNR varies across different parameters. Similar trends persist across the other PE results presented here as well as in the subsequent discussion.

In Fig.~\ref{fig:Theta_LS_30_Theta_JN_0_XPNR_mode_cpmrsn} we examine whether the biases can be ascribed to poor modelling of the higher modes. Here we restrict ourselves to the {\tt XPNR} model (the only model calibrated to precessing NR simulations among the three models considered) and the $\theta_{\rm JN} = 0^\circ$ signal, and recover a signal that contains only the $\ell = 2$ modes with a corresponding $\ell=2$ model, and compare the results with the recovery of the full signal with the full model. We find that restricting to $\ell=2$ modes broadens the posteriors, as we might expect with a signal lacking rich higher-mode structure, but we also see that the bias is now \emph{worse}. This illustrates that the models are of poor accuracy even at the level of only the $\ell=2$ modes.

In Fig.~\ref{fig:Theta_LS_150_Theta_JN_0_80} we consider the 
$\theta_{\rm LS_1} = 150^\circ$ system (NR simulation ${\tt CF\_{85}}$ of Table~\ref{table:metadata}, with orientations $\theta_{\rm JN} = 0^\circ$ and $\theta_{\rm JN} = 80^\circ$. This configuration exhibits far stronger precession, but, due to the large anti-aligned spin components, merges at a low frequency and includes only a small number of GW cycles in the frequency range of our analysis (see Figs.~\ref{fig:NR_sim} and \ref{fig:hp_cmprsn}). Despite the small number of measureable GW cycles and the relatively good agreement by eye in Fig.~\ref{fig:hp_cmprsn}, we still see large biasses with all models for at least one orientation. Now the error in the recovery of the secondary mass can be as large as 100\%, and the mass ratio can be recovered below $q=10$. We note that, although in all cases the mismatches were not a useful indicator of the level of bias, the mismatches do allow us to correctly predict that the most accurate model for this configuration is {\tt TPHM}.

We note that the posteriors for other parameters are also biased, although the magnitude of the bias varies across parameters. The mass and spin parameters shown here are neither the only nor necessarily the most or least affected parameters; they are presented simply as representative examples of the observed parameter biases.

Finally, we aim to understand how NR waveform errors map onto the waveform model manifold, i.e., how sensitive waveform models are to the numerical errors of NR simulations in the parameter estimation process.
For this purpose, we generate three synthetic GW signals from the NR waveforms of configuration ${\tt CF\_{81}}$ in Table~\ref{table:metadata}, corresponding to varying numerical resolutions $N=120, 144, 172$, at a redshifted total mass of $150M_{\odot}$ and $\theta_{\text{JN}} = 48^\circ$, choosing the reference phase so that $\theta_{\text{LN}} = 60^\circ$.
We then perform parameter estimation on these signals using the {\tt XPNR} waveform model.
The posterior distributions of the component masses and the primary spin components $\vec{\chi}_{1}^{\parallel}$ and $\vec{\chi}_{1}^{ \perp}$ at the reference frequency are shown in Fig.~\ref{fig:Theta_LS_30_Mtot_150_Theta_JN_48_XPNR_N_120_144_172}.

We see that the posteriors are only slightly shifted for the signals corresponding to the three NR resolutions, and the $N=144$ and $N=172$ results are almost identical. This is in stark contrast to what we might expect from the mismatches between the NR waveforms, which suggest that the $N=120$ configuration would be distinguishable from the $N=172$ configuration at an SNR of only $\sim$9. These results suggest that the error in the NR waveforms are predominantly orthogonal to the physical signal manifold, and do not incur parameter biases. This may have implications for the NR accuracy necessary for model construction, which we will return to in the discussion.

We note that we did not consider all current models in this study, for example the {\tt TEOBResumS} models~\cite{Nagar:2018zoe, Akcay:2020qrj, Gamba:2021ydi}.
However, although their construction differs from the {\tt XPNR}, {\tt TPHM} and {\tt v5PHM} models, they do not include any additional information: they use the same inspiral ingredients from post-Newtonian theory or the effective-one-body framework, employ similar techniques to treat the merger and ringdown, and also lack calibration to precessing NR waveforms.
We therefore expect that our conclusions apply to all current models: they are not suitable for GW astronomy involving high-mass-ratio, high-spin binaries with significant precession.
This limitation may not be an urgent problem for current detectors, where large mass ratios and high spins are both rare, and we expect better-tuned models in the near future, assisted by these NR waveforms. Nevertheless, these results starkly illustrate that models cannot be trusted far outside their region of calibration and validation.

\section{Discussion}\label{sec:conclu}
We have presented a series of five NR simulations of black-hole binaries with a high mass-ratio ($q=18$) and significant precession. These extend the catalog presented in Ref.~\cite{Hamilton:2023qkv}, and are a first step in filling a significant gap in simulation parameter-space coverage. In each simulation the larger black hole has a dimensionless spin of 0.8, and the simulations span five spin misalignment angles. The simulations cover $\sim$3000\,$M$ before merger (between 30 and 49 GW cycles), which make them ideal for extending current precessing models into the high-mass-ratio regime, and for assessing the accuracy of current models. 

Estimates of the accuracy of the NR waveforms, and comparisons (both through mismatches and parameter-estimation examples) allow us to draw a number of conclusions, and raise several important points. 

We preface our conclusions with the caveat that our statements are restricted to high-mass binaries, where the late inspiral and merger-ringdown constitute significant power in the signal. The length of our NR simulations preclude statements about binaries with lower masses. However, we note that if we assume a minimum black-hole mass of 5\,$M_\odot$, then the lowest total mass for a $q=18$ binary will be 95\,$M_\odot$, and masses much lower than we have considered are unlikely. 

Our main conclusion is that current state-of-the-art waveform models cannot be trusted in measurements of high-mass-ratio high-precession binaries. Across a set of four signals that would be observed with SNR $\rho = 50$ in the Advanced LIGO detector at design sensitivity, we typically find that even the most basic property, the masses of the binaries, are recovered with errors up to $\sim$100\%; a mass-ratio 18 system can be measured to have a mass ratio of less than 10. (See Fig.~\ref{fig:Theta_LS_150_Theta_JN_0_80} for the most striking example.) No model is consistently more or less accurate than any other. The primary spin magnitude shows biases where the true value is up to 20\% away from the 90\% credible region of the measurement, and the lower bound on the measured spin magnitude is half of the true value. The spin orientation can be biased by up to 20$^\circ$, although this is sufficient to determine if the spin is predominantly aligned or anti-aligned with the binary's orbital angular momentum. 

Although these results are striking, they should perhaps not be surprising. Of the three models we considered, precession dynamics are tuned to NR simulations only in {\tt XPNR}, and even in that case only up to $q=8$, and only for the dominant multipole in the co-precessing frame. We have also seen in earlier work that current models show significant disagreement at high mass ratios, even in the non-precessing sector~\cite{Pompili:2023tna}. As a precessing-binary example, Ref.~\cite{Dhani:2024jja} presents detailed results for a highly precessing binary with mass-ratio $q=10$, with mass-measurement errors of up to 10\%. In that example the binary had a total mass of $\sim$25\,$M_\odot$, so the signal was dominated by the inspiral, where we might expect current models (all based on PN and EOB inspiral results) to be in better agreement. Based on those results, we should not be surprised that systems with higher mass (i.e., more signal power in the poorly modeled merger regime), and higher mass ratio would incur even larger biases. However, the study in Ref.~\cite{Dhani:2024jja} did not involve NR simulations, and could therefore only consider disagreements between models, and could not make absolute statements about the biases that would be observed from a true BBH signal. In this work we \emph{have} used fully general-relativistic NR results, and have been able to conclusively show that current models are almost entirely inadequate for measurements of high-mass-ratio highly precessing binaries. 

Of the $\sim$200 binaries detected to date in LVK observations, the largest well-measured mass ratio is $q=10$~\cite{GW190814}, so we can expect larger mass ratios to be extremely rare with current detectors, and high-mass-ratio binaries with high spins are expected to be even rarer. However, intermediate-mass-ratio binaries are an important source for next-generation ground-based detectors and the space-based detector LISA, and for those observations we need far more accurate models. And to produce more accurate models, we need high-mass-ratio NR simulations. 

This raises the question of NR accuracy requirements and computational cost. There is still no robust method to determine minimal NR or model accuracy requirements for GW astronomy. Our new NR waveforms provide a striking illustration of this issue. The mismatch errors for our base-resolution simulation are very high ($\sim$0.04), suggesting that the lowest and highest resolution waveforms would be distinguishable in an observation at SNRs above $\sim$5. In PE examples, however, we find the three NR resolutions are \emph{not} clearly distinguishable, even at an SNR of 50. (See Fig.~\ref{fig:Theta_LS_30_Mtot_150_Theta_JN_48_XPNR_N_120_144_172}.) This illustrates the degree to which NR errors can be orthogonal to waveform variations with respect to physical parameters. 

Let us consider one way in which this could happen, to illustrate how NR errors may be orthogonal to measurement biases. The phase error in an NR simulations increases more rapidly as the binary approaches merger; as the black holes orbit more rapidly the numerical error per timestep increases. A slower or faster inspiral will be partially degenerate with a change in total mass (which is an overall scaling in the time), and with mass ratio and aligned spins, where the changes in inspiral rate are frequency dependent~\cite{Baird:2012cu}. However, after merger the signal has a fixed ringdown frequency, and the even low-resolution simulations tend to capture this correctly, so that the phase error plateaus after merger. This change is \emph{not} degenerate with a change in total mass or mass ratio, since these will change the final mass and spin, and therefore the ringdown frequency.
In a signal with comparable power in both the inspiral and merger-ringdown, we may find that, although the NR waveform has the wrong phasing, the phase errors are not clearly degenerate with a biased mass and spins; in a search or parameter-estimation exercise such a signal may be recovered with a lower SNR, but not necessarily with a significant bias in the masses and spins. This illustrates the difficulty in specifying the NR accuracy requirements for future detectors. Our results suggests that, while ideally more accurate NR waveforms than those presented here will ultimately be necessary to calibrate waveform models, these first simulations may be sufficient to produce models that will be reliable for signals of moderate SNRs ($<100$). The full determination of NR and waveform accuracy requirements remains an open question.

We now turn to computational cost. The most accurate simulation we produced as part of our convergence series, the $N=172$ simulation of the $\theta_{\rm LS} = 30^\circ$ configuration (with a mismatch error of 0.007), is still less accurate than typical simulations in the earlier BAM catalog~\cite{Hamilton:2023qkv} (0.004), yet the computational cost is far higher. The $N=172$ simulation alone cost 13M core hours, while the \emph{entire catalog} of 80 simulations in Ref.~\cite{Hamilton:2023qkv} cost 25M core hours. This illustrates the well-known problem that, even though high-mass-ratio simulations do not present any known technical challenge (e.g., proof-of-principle moving-puncture simulations have been performed at mass ratios up to $\sim$1000~\cite{Lousto:2022hoq}), binary simulations that include a sufficient number of orbits before merger are prohibitively expensive with current codes and methods. If the NR community is to produce accurate NR simulations in the intermediate-mass-ratio regime ($10 < q < 1000$) then improved codes and methods are crucial. In addition it is also crucial that we reach a better understanding of the relationship between NR accuracy and GW measurement accuracy.

\vspace*{\baselineskip}
\section*{Acknowledgements}\label{sec:acknowledgements}
The authors thank Jannik Mielke for useful comments as the LIGO-Virgo-KAGRA internal reviewer, and Lorenzo Pompili and Carlos Lousto for their comments.
P.M. and M.H. acknowledge the Science and Technology Facilities Council (STFC) for support through Grants ST/V005618/1 and UKRI2489. J.T. acknowledges support from the NASA LISA Preparatory Science grant 20-LPS20-0005. This research used the supercomputing facilities at Cardiff University operated by Advanced Research Computing at Cardiff 
(ARCCA) on behalf of the Cardiff Supercomputing Facility and the HPC Wales and Supercomputing Wales (SCW) projects. 
We acknowledge the support of the latter, which is part-funded by the European Regional Development Fund (ERDF) via the Welsh Government. 
The authors are also grateful for computational resources provided by the Cardiff University and support by STFC Grants ST/I006285/1, ST/V005618/1 and UKRI2489. The authors are grateful for computational resources provided by the LIGO Laboratory and supported by National Science Foundation Grants No. PHY-0757058 and No.
PHY-0823459. This manuscript has the LIGO preprint No. P2600112.

Various plots and analyses in this paper were made using Python software packages \texttt{LALSuite}~\cite{lalsuite}, \texttt{PyCBC}~\cite{alex_nitz_2024_10473621}, PESummary~\cite{Hoy:2020vys}, \texttt{Matplotlib}~\cite{Hunter:2007}, \texttt{Numpy}~\cite{harris2020array}, and \texttt{Scipy}~\cite{2020SciPy-NMeth}.

\section*{DATA AVAILABILITY}
The data that support the findings of this article are not publicly available. The data are available from the authors upon reasonable request.

\bibliography{refs}

\clearpage

\onecolumngrid
\appendix*
\section{Supplemental Materials}
\begin{figure*}[t]
    \centering
    \includegraphics[width=\textwidth]{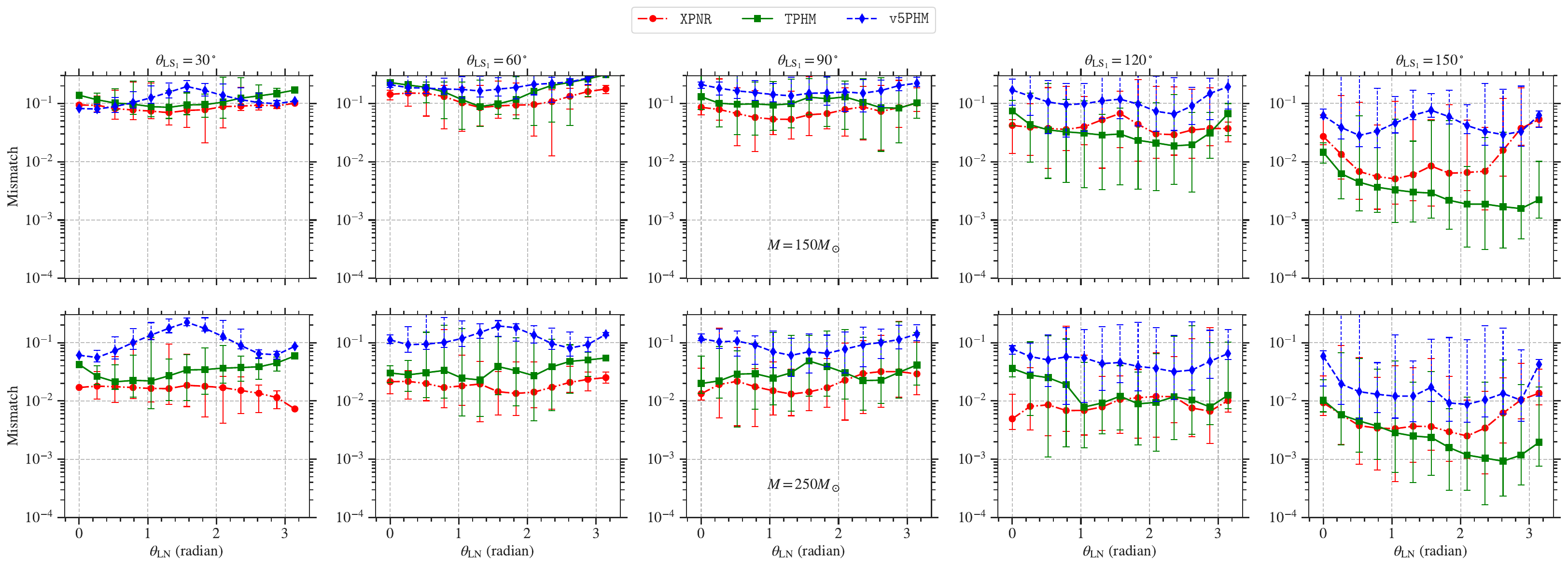}
    \caption{Mismatch between the dominant quadrupolar $(\ell, |m|) = (2, 2)$ modes of five reported NR simulations and the corresponding modes of three state-of-the-art waveform models (shown in the legend) as a function of the binary inclination angle---quantified by the angle between the line of sight and the Newtonian orbital angular momentum--at redshifted total masses of $150 M_{\odot}$ (top row) and $250 M_{\odot}$ (bottom row). The mismatch shown here is optimized over luminosity distance, coalescence time and phase, sky location, polarization, and precession angle. The reported NR simulations represent high mass-ratio, precessing BBH systems with a mass ratio of 18 and a dimensionless spin magnitude of 0.8 on the larger BH (the smaller BH being nonspinning), and the five simulations correspond to spin misalignment angles of $30^\circ$, $60^\circ$, $90^\circ$, $120^\circ$, and $150^\circ$, with each column in the figure representing one of these angles. Each solid marker shows the SNR-weighted mismatch, and the horizontal lines indicate the range from minimum to maximum mismatch across the considered signal polarizations and phases.}
    \label{fig:l2mismatches}
\end{figure*}

Sec.~IV and Fig.~5 of the main text present only the minimum and maximum values, and the overall range, of optimized mismatches between the waveform models and the NR waveforms. This Supplemental Material provides additional mismatch results that complement the main text. In particular, we present the optimized mismatch as a function of the binary orbital inclination angle $\theta_{\rm LN}$, for two redshifted total masses, $M=150M_{\odot}$ and $250M_{\odot}$. 

We first focus on the dominant quadrupolar $\ell = 2$ modes in both the waveform models and the NR simulations, and compute the corresponding optimized mismatches as described in the main text (see Sec.~II A and ~IV for more details). Fig.~\ref{fig:l2mismatches} presents the mismatches as a function of $\theta_{\rm LN}$, for the two redshifted total masses mentioned above. It employs the five precessing NR simulations introduced in the main text, each representing a different spin misalignment configuration. Solid markers indicate the SNR-weighted mismatch, while horizontal lines denote the range of mismatches across the sampled polarizations and phases.

We find that, for the first three considered values of $\theta_{\text{LS}_1}$, the SNR–weighted mismatches are typically greater than $10^{-2}$, corresponding to a bias SNR of $\ge 12$ for single-parameter measurements, according to the conservative mismatch criterion ($\mathfrak{M}\le1.35/\rho^{2}$) mentioned above. The lowest mismatch is found to be $3 \times 10^{-3}$, corresponding to a bias SNR of $21$. For the latter two values of $\theta_{\text{LS}_1}$, the SNR–weighted mismatches are generally above $10^{-3}$, corresponding to a bias SNR of $\ge 37$. 
The mismatches are typically smaller for $\theta_{\text{LS}_1}=150^\circ$ compared to $\theta_{\text{LS}_1}=30^\circ$, and for $\theta_{\text{LS}_1}=120^\circ$ compared to $\theta_{\text{LS}_1}=60^\circ$. This occurs because the number of cycles within a given frequency band decreases significantly as the spin–orbit misalignment angle increases due to the hang-up effect. We also find that $M= 150 M_{\odot}$ systems generally yield worse mismatch compared to $250 M_{\odot}$ cases, because lower-mass systems will have more GW cycles in the LIGO frequency band.

\begin{figure*}[t]
    \centering
    \includegraphics[width=\textwidth]{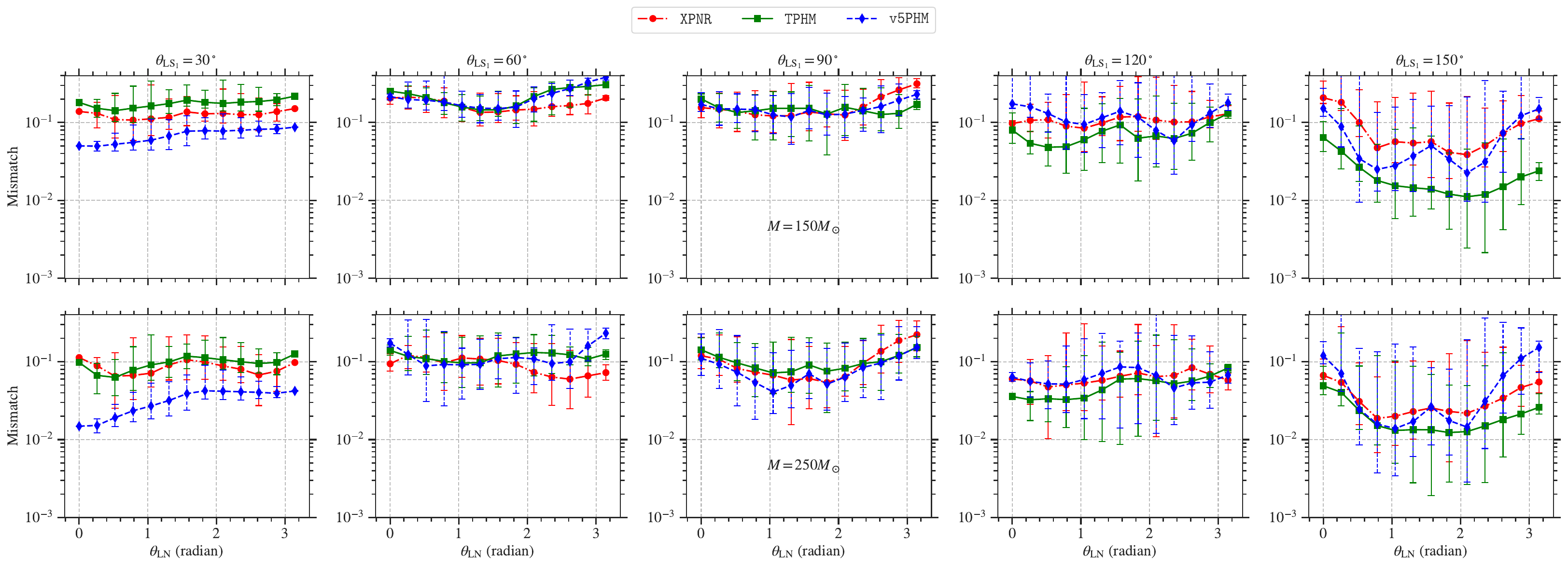}
    \caption{Mismatch between NR simulations and waveform models including all available modes, analogous to Figure~\ref{fig:l2mismatches} but extending beyond the dominant quadrupolar modes.}
    \label{fig:allmodemismatches}
\end{figure*}

Next, we focus on all the available modes in the waveform models and on modes up to $\ell=4$ in the NR simulations, and compute the corresponding mismatches. We have not considered the memory modes (modes with $m=0$) from the NR simulations (the available modes in the three waveform models are listed in Sec.~I of the main text).
We present the results in Fig.~\ref{fig:allmodemismatches}. By comparing the results shown in Fig.~\ref{fig:l2mismatches} and Fig.~\ref{fig:allmodemismatches}, we find that the mismatches between the waveform models and the NR simulations become larger when higher-order modes are included in addition to the dominant quadrupolar modes. Here, we find that SNR–weighted mismatches are always greater than $10^{-2}$, corresponding to a bias SNR of $\ge 12$, while the lowest mismatch is $2 \times 10^{-3}$, corresponding to a bias SNR of $26$ for single-parameter measurements, according to the conservative mismatch criterion. All the adopted waveform models perform roughly equally poorly in terms of mismatch, and none can be considered superior to the others. All other trends remain largely similar to those shown in Fig.~\ref{fig:l2mismatches}. Current models require significant improvement in the region of parameter space covered by the reported NR simulations.

\end{document}

%% file: metadata-tabular-q18.tex
\begin{tabular*}{\textwidth}{@{\extracolsep{\fill}} ll l l l l l l l l l l l l }
\toprule[0.16em]
\addlinespace[0.25em]
Name & $q$ & $\chi_1$ & $\theta_{\mathrm{LS}_{1}} \left(^\circ\right)$ & $\chi_{\mathrm{eff}}$ & $\chi_{\rm p}$ & $D/M$ & $e$ & $M\omega_{\mathrm{orb}}$ & $t_{\rm M}/M$ & $N_{\mathrm{orb}}$ & $M_{f}$ & $\chi_{f}$ & $V_{\rm kick}$ \\
& & & & & & & \scriptsize{($\times 10^{-3}$)} & \scriptsize{($\times 10^{-2}$)} & & & & & \scriptsize{(km s$^{-1}$)} \\
\addlinespace[0.25em]
\hline\hline
\addlinespace[0.5em]
\midrule
\texttt{CF\_81} & 18 & 0.80 & 30 & 0.66 & 0.40 & 8.33 & 1.2 & 3.4 & 3074 & 24.6 & 0.990 & 0.814 & 53\\
\addlinespace[0.5em]
\texttt{CF\_82} & 18 & 0.80 & 60 & 0.38 & 0.69 & 8.93 & 1.5 & 3.2 & 3227 & 23.4 & 0.992 & 0.787 & 148\\
\addlinespace[0.5em]
\texttt{CF\_83} & 18 & 0.80 & 90 & 0.00 & 0.80 & 9.70 & 2.0 & 2.9 & 3265 & 20.4 & 0.994 & 0.729 & 171\\
\addlinespace[0.5em]
\texttt{CF\_84} & 18 & 0.80 & 120 & $-0.38$ & 0.69 & 10.47 & 1.3 & 2.6 & 3150 & 17.3 & 0.995 & 0.645 & 117\\
\addlinespace[0.5em]
\texttt{CF\_85} & 18 & 0.80 & 150 & $-0.66$ & 0.40 & 11.05 & 2.4 & 2.5 & 3014 & 15.2 & 0.996 & 0.561 & 34\\
\addlinespace[0.5em]
\bottomrule[0.16em]
\end{tabular*}